\documentclass[10pt,preprint]{aastex}

\newcommand{\avg}[1]{{\langle{#1}\rangle}}
\newcommand{\Avg}[1]{{\left\langle{#1}\right\rangle}}
\def\simless{\mathbin{\lower 3pt\hbox
	{$\,\rlap{\raise 5pt\hbox{$\char'074$}}\mathchar"7218\,$}}} 
\def\simgreat{\mathbin{\lower 3pt\hbox
	{$\,\rlap{\raise 5pt\hbox{$\char'076$}}\mathchar"7218\,$}}} 

\newcommand{\petroratio}{{{\mathcal{R}}_P}}
\newcommand{\petroradius}{{{r}_P}}
\newcommand{\petronumber}{{{N}_P}}
\newcommand{\petroratiolim}{{{\mathcal{R}}_{P,\mathrm{lim}}}}

\newcounter{thefigs}
\newcommand{\fignum}{\arabic{thefigs}}

\newcounter{thetabs}

\newcounter{address}

\slugcomment{Submitted to \aj}

\shortauthors{Blanton {\it et al.} (2000)}
\shorttitle{Luminosity Function in SDSS Commissioning Data}

\begin{document}
 

\title{The Luminosity Function of Galaxies in SDSS Commissioning
Data$^1$}



\author{
Michael R. Blanton\altaffilmark{\ref{Fermilab}},  
Julianne Dalcanton\altaffilmark{\ref{UW}},
Daniel Eisenstein\altaffilmark{\ref{Chicago},\ref{Hubble}},
Jon Loveday\altaffilmark{\ref{Sussex}},
Michael A. Strauss\altaffilmark{\ref{Princeton}},
Mark SubbaRao\altaffilmark{\ref{Chicago}},
David H. Weinberg\altaffilmark{\ref{Ohio}},
John E. Anderson, Jr.\altaffilmark{\ref{Fermilab}},
James Annis\altaffilmark{\ref{Fermilab}},
Neta A. Bahcall\altaffilmark{\ref{Princeton}},
Mariangela Bernardi\altaffilmark{\ref{Chicago}},
J. Brinkmann\altaffilmark{\ref{APO}},
Robert J. Brunner\altaffilmark{\ref{Caltech}},
Scott Burles\altaffilmark{\ref{Fermilab}},
Larry Carey\altaffilmark{\ref{UW}},
Francisco J. Castander\altaffilmark{\ref{Chicago}, \ref{Pyrenees}},
Andrew J. Connolly\altaffilmark{\ref{Pitt}},
Istv\'an Csabai\altaffilmark{\ref{JHU}},
Mamoru Doi\altaffilmark{\ref{Tokyo}},
Douglas Finkbeiner\altaffilmark{\ref{Berkeley}},
Scott Friedman\altaffilmark{\ref{JHU}},
Joshua A. Frieman\altaffilmark{\ref{Fermilab}},
Masataka Fukugita\altaffilmark{\ref{CosmicRay},\ref{IAS}},
James E. Gunn\altaffilmark{\ref{Princeton}},
G. S. Hennessy\altaffilmark{\ref{USNO}},
Robert B. Hindsley\altaffilmark{\ref{USNO}},
David W. Hogg\altaffilmark{\ref{Princeton}},
Takashi Ichikawa\altaffilmark{\ref{Tokyo}},
\v{Z}eljko Ivezi\'{c}\altaffilmark{\ref{Princeton}},
Stephen Kent\altaffilmark{\ref{Fermilab}},
G. R.~Knapp\altaffilmark{\ref{Princeton}},
D. Q.~Lamb\altaffilmark{\ref{Chicago}},
R. French Leger\altaffilmark{\ref{UW}},
Daniel C. Long\altaffilmark{\ref{APO}},
Robert H. Lupton\altaffilmark{\ref{Princeton}},
Timothy A.~McKay\altaffilmark{\ref{Michigan}},
Avery Meiksin\altaffilmark{\ref{Edinburgh}},
Aronne Merelli\altaffilmark{\ref{Caltech}},
Jeffrey A. Munn\altaffilmark{\ref{USNO}},
Vijay Narayanan\altaffilmark{\ref{Princeton}},
Matt Newcomb\altaffilmark{\ref{CarnegieMellon}},
R. C. Nichol\altaffilmark{\ref{CarnegieMellon}},
Sadanori Okamura\altaffilmark{\ref{Tokyo}},
Russell Owen\altaffilmark{\ref{UW}},
Jeffrey R.~Pier\altaffilmark{\ref{USNO}},
Adrian Pope\altaffilmark{\ref{JHU}},
Marc Postman\altaffilmark{\ref{STScI}},
Thomas Quinn\altaffilmark{\ref{UW}},
Constance M. Rockosi\altaffilmark{\ref{Chicago}},
David J. Schlegel\altaffilmark{\ref{Princeton}},
Donald P. Schneider\altaffilmark{\ref{PennState}}, 
Kazuhiro Shimasaku\altaffilmark{\ref{Tokyo}},
Walter A. Siegmund\altaffilmark{\ref{UW}},
Stephen Smee\altaffilmark{\ref{Maryland}},
Yehuda Snir\altaffilmark{\ref{CarnegieMellon}},
Chris Stoughton\altaffilmark{\ref{Fermilab}},
Christopher Stubbs\altaffilmark{\ref{UW}},
Alexander S.~Szalay\altaffilmark{\ref{JHU}},
Gyula P.~Szokoly\altaffilmark{\ref{Potsdam}},
Aniruddha R.~Thakar\altaffilmark{\ref{JHU}},
Christy Tremonti\altaffilmark{\ref{JHU}},
Douglas L. Tucker\altaffilmark{\ref{Fermilab}},
Alan Uomoto\altaffilmark{\ref{JHU}},
Dan vanden Berk\altaffilmark{\ref{Fermilab}},
Michael S. Vogeley\altaffilmark{\ref{Drexel}},
Patrick Waddell\altaffilmark{\ref{UW}},
Brian Yanny\altaffilmark{\ref{Fermilab}},
Naoki Yasuda\altaffilmark{\ref{NAOJ}},
and Donald G.~York\altaffilmark{\ref{Chicago}}
}

\altaffiltext{1}{Based on observations obtained with the
Sloan Digital Sky Survey} 
\setcounter{address}{2}
\altaffiltext{\theaddress}{
\stepcounter{address}
Fermi National Accelerator Laboratory, P.O. Box 500,
Batavia, IL 60510
\label{Fermilab}}
\altaffiltext{\theaddress}{
\stepcounter{address}
Department of Astronomy, University of Washington,
Box 351580,
Seattle, WA 98195 
\label{UW}}
\altaffiltext{\theaddress}{
\stepcounter{address}
University of Chicago, Astronomy \&
Astrophysics Center, 5640 S. Ellis Ave., Chicago, IL 60637
\label{Chicago}}
\altaffiltext{\theaddress}{
\stepcounter{address}
Hubble Fellow 
\label{Hubble}}
\altaffiltext{\theaddress}{
\stepcounter{address}
Sussex Astronomy Centre,
University of Sussex,
Falmer, Brighton BN1 9QJ, UK
\label{Sussex}}
\altaffiltext{\theaddress}{
\stepcounter{address}
Princeton University Observatory, Princeton,
NJ 08544
\label{Princeton}}
\altaffiltext{\theaddress}{
\stepcounter{address}
Ohio State University,
Department of Astronomy,
Columbus, OH 43210
\label{Ohio}}
\altaffiltext{\theaddress}{
\stepcounter{address}
Apache Point Observatory,
2001 Apache Point Road,
P.O. Box 59, Sunspot, NM 88349-0059
\label{APO}}
\altaffiltext{\theaddress}{
\stepcounter{address}
Department of Astronomy, California Institute of Technology,
Pasadena, CA 91125
\label{Caltech}}
\altaffiltext{\theaddress}{
\stepcounter{address}
Observatoire Midi-Pyr\'en\'ees, 
14 ave Edouard Belin, Toulouse, F-31400, France
\label{Pyrenees}}
\altaffiltext{\theaddress}{
\stepcounter{address}
University of Pittsburgh,
Department of Physics and Astronomy,
3941 O'Hara Street,
Pittsburgh, PA 15260
\label{Pitt}}
\altaffiltext{\theaddress}{
\stepcounter{address}
Department of Physics and Astronomy, The Johns Hopkins University,
Baltimore, MD 21218
\label{JHU}}
\altaffiltext{\theaddress}{
\stepcounter{address}
Department of Astronomy and Research Center for 
the Early Universe,
School of Science, University of Tokyo,
Tokyo 113-0033, Japan
\label{Tokyo}}
\altaffiltext{\theaddress}{
\stepcounter{address}
UC Berkeley, Dept. of Astronomy, 601 Campbell Hall, Berkeley, CA  94720-3411
\label{Berkeley}}
\altaffiltext{\theaddress}{
\stepcounter{address}
Institute for Cosmic Ray Research, University of
Tokyo, Midori, Tanashi, Tokyo 188-8502, Japan
\label{CosmicRay}}
\altaffiltext{\theaddress}{
\stepcounter{address}
Institute for Advanced Study, Olden Lane,
Princeton, NJ 08540
\label{IAS}}
\altaffiltext{\theaddress}{
\stepcounter{address}
U.S. Naval Observatory,
3450 Massachusetts Ave., NW,
Washington, DC  20392-5420
\label{USNO}}
\altaffiltext{\theaddress}{
\stepcounter{address}
University of Michigan, Department of Physics,
500 East University, Ann Arbor, MI 48109
\label{Michigan}}
\altaffiltext{\theaddress}{
\stepcounter{address}
Department of Physics \& Astronomy,
The University of Edinburgh,
James Clerk Maxwell Building,
The King's Buildings,
Mayfield Road,
Edinburgh EH9 3JZ, UK
\label{Edinburgh}}
\altaffiltext{\theaddress}{
\stepcounter{address}
Department of Physics, Carnegie Mellon University, 
5000 Forbes Avenue, Pittsburgh, PA 15213-3890 
\label{CarnegieMellon}}
\altaffiltext{\theaddress}{
\stepcounter{address}
Space Telescope Science Institute, Baltimore, MD 21218
\label{STScI}}
\altaffiltext{\theaddress}{
\stepcounter{address}
Department of Astronomy and Astrophysics,
The Pennsylvania State University,
University Park, PA 16802
\label{PennState}}
\altaffiltext{\theaddress}{
\stepcounter{address}
Department of Astronomy,
University of Maryland,
College Park, MD 20742-2421 
\label{Maryland}}
\altaffiltext{\theaddress}{
\stepcounter{address}
Astrophysikalisches Institut Potsdam,
An der Sternwarte 16, D-14482 Potsdam, Germany
\label{Potsdam}}
\altaffiltext{\theaddress}{
\stepcounter{address}
Department of Physics, Drexel University, Philadelphia, PA 19104
\label{Drexel}}
\altaffiltext{\theaddress}{
\stepcounter{address}
National Astronomical Observatory, Mitaka, Tokyo 181-8588, Japan
\label{NAOJ}}

\clearpage

\begin{abstract}
In the course of its commissioning observations, the Sloan Digital Sky
Survey (SDSS) has produced one of the largest redshift samples of
galaxies selected from CCD images.  Using 11,275 galaxies complete to
$r^\ast = 17.6$ over 140 square degrees, we compute the luminosity
function of galaxies in the $r^\ast$ band over a range $-23 <
M_{r^\ast} < -16$ (for $h=1$). The result is well-described by a
Schechter function with parameters $\phi_\ast= (1.46\pm 0.12) \times
10^{-2}$ $h^3$ Mpc$^{-3}$, $M_\ast = -20.83\pm 0.03$, and $\alpha =
-1.20\pm 0.03$. The implied luminosity density in $r^\ast$ is
$j\approx (2.6\pm 0.3) \times 10^{8} h L_\odot$Mpc$^{-3}$. We find
that the surface brightness selection threshold has a negligible
impact for $M_{r^\ast}<-18$.  Using subsets of the data, we measure
the luminosity function in the $u^\ast$, $g^\ast$, $i^\ast$, and
$z^\ast$ bands as well; the slope at low luminosities ranges from
$\alpha=-1.35$ to $\alpha=-1.2$. We measure the bivariate distribution
of $r^\ast$ luminosity with half-light surface brightness, intrinsic
$g^\ast-r^\ast$ color, and morphology. In agreement with previous
studies, we find that high surface brightness, red, highly
concentrated galaxies are on average more luminous than low surface
brightness, blue, less concentrated galaxies. An important feature of
the SDSS luminosity function is the use of Petrosian magnitudes, which
measure a constant fraction of a galaxy's total light regardless of
the amplitude of its surface brightness profile. If we synthesize
results for $R_{\mathrm{GKC}}$-band or $b_j$-band using these
Petrosian magnitudes, we obtain luminosity densities 2 times that
found by the Las Campanas Redshift Survey in $R_{\mathrm{GKC}}$ and
1.4 times that found by the Two-degree Field Galaxy Redshift Survey in
$b_j$. However, we are able to reproduce the luminosity functions
obtained by these surveys if we also mimic their isophotal limits for
defining galaxy magnitudes, which are shallower and more redshift
dependent than the Petrosian magnitudes used by the SDSS.
\end{abstract}


%
%

\section{Motivation}
\label{motivation}

A fundamental characteristic of the galaxy population, which has been
the subject of study at least since \citet{hubble36b}, is the
distribution of their luminosities.  Given the broad spectral energy
distributions of galaxies (especially in the presence of dust), the
bolometric luminosity of galaxies is currently too difficult to
measure to meaningfully study, so we must be satisfied with observing
the luminosity function in some wavelength bandpass. In this paper we
measure the luminosity function of local ($z<0.2$) galaxies in five
optical bandpasses, between about 3000 \AA\ and 10000 \AA. We also
investigate the correlation of luminosity with other galaxy
properties, such as surface brightness, intrinsic color, and
morphology. This quantitative characterization of the local galaxy
population provides the basic data that a theory of galaxy formation
must account for and an essential baseline for studies of galaxy
evolution at higher redshifts.

The most recent determinations of the optical luminosity function of
``field'' galaxies (those selected without regard to local density)
have come from large flux-limited redshift surveys. They include the
luminosity function measurements of \citet{lin96a}, using the Las
Campanas Redshift Survey (LCRS; \citealt{shectman96a}) in the LCRS
$R_{\mathrm{GKC}}$-band (around 6500 \AA), and of \citet{folkes99a}
using the Two-degree Field Galaxy Redshift Survey (2dFGRS;
\citealt{colless99a}) in the $b_j$-band (around 4500 \AA).  Other
recent determinations of the local optical luminosity function have
been mostly in the $B$- or $b_j$-bands, for example the Nearby Optical
Galaxy sample (\citealt{marinoni99a}), the Optical Redshift Survey
(\citealt{santiago96a}), the Stromlo-APM Redshift Survey
(\citealt{loveday92a}), the Durham/UKST Galaxy Redshift survey
(\citealt{ratcliffe98a}), and the ESO Slice Project Galaxy Redshift
Survey (\citealt{zucca97a}). Of particular interest here is the
ability to use the multi-band photometry of the SDSS to {\it
meaningfully} compare our results to those of these other surveys.




In this paper, we use imaging and spectroscopy of 11,275 galaxies over
about 140 square degrees complete to $r^\ast = 17.6$ from the Sloan
Digital Sky Survey (SDSS; \citealt{york00a}) commissioning data to
calculate the luminosity function of galaxies. In Section \ref{data}
we describe the source catalog and the redshift sample, with
particular attention to our definition of galaxy magnitudes, which is
based on a modified form of the
\citet{petrosian76a} system. In Section \ref{method}, we describe the
maximum-likelihood method used to calculate the luminosity function. In
Section \ref{results}, we present the results for the luminosity
function in each band, for the luminosity density of the universe, and
for the dependence of galaxy luminosity on surface brightness, color and
morphology. In Section \ref{isopetro}, we examine the effects of using
Petrosian magnitudes rather than isophotal magnitudes for the luminosity
function. In Section \ref{others}, we compare our results to those of
Lin {\it et al.} (1996; LCRS) and Folkes {\it et al.} (1999;
2dFGRS). We show that the SDSS luminosity function implies a
substantially higher mean luminosity density than either of these
previous measurements, but that we can reproduce the results of both
surveys if we mimic their isophotal definitions of galaxy magnitudes.  We
conclude in Section \ref{conclusions}.

\section{SDSS Commissioning Data}
\label{data}

\subsection{Description of the Survey}

The SDSS (\citealt{york00a}) will produce imaging and spectroscopic
surveys over $\pi$ steradians in the Northern Galactic Cap. A
dedicated 2.5m telescope (Siegmund {\it et al.}, in preparation) at
Apache Point Observatory, Sunspot, New Mexico, will image the sky in
five bands ($u'$, $g'$, $r'$, $i'$, $z'$; centered at 3540 \AA, 4770
\AA, 6230
\AA, 7630 \AA, and 9130 \AA\ respectively; \citealt{fukugita96a}) using
a drift-scanning, mosaic CCD camera (\citealt{gunn98a}), detecting
objects to a flux limit of $r'\sim 23$. Approximately 900,000
galaxies, (down to $r_{\mathrm{lim}}'\approx 17.65$), 100,000 Bright
Red Galaxies (BRGs; Eisenstein {\it et al.}, in preparation), and
100,000 QSOs (\citealt{fan99a}; Newberg {\it et al.}, in preparation)
will be targeted for spectroscopic follow up using two digital
spectrographs on the same telescope. The survey has completed its
commissioning phase, during which the data analyzed here were
obtained.

As of January 2001, the SDSS has imaged around 2,000 square degrees
of sky and taken spectra of approximately 100,000 objects. In this
paper, we concentrate only on $\sim$ 230 square degrees along the
Celestial Equator in the region bounded by $145^\circ<\alpha
<236^\circ$ and $-1.25^\circ<\delta<1.25^\circ$ (J2000). This region
was imaged during two runs, known as SDSS runs 752 and 756, over two
nights in March 1999. Each run consists of six columns of data, each
slightly wider than 0.2 degrees and separated by about the same
amount; the runs are interleaved to form a complete, 2.5 degree wide,
stripe. The seeing varied over the course of the runs from about
$1.2''$ to $2''$, with a median of approximately $1.5''$. In this
area, the spectroscopic survey has so far obtained more than 11,275
redshifts of galaxies selected from these runs, over about 140 square
degrees of sky, from which we here calculate the luminosity function.

\subsection{Identifying and Measuring Galaxies}
\label{photo}

The images produced by the camera are analyzed by a photometric
pipeline specifically written for reducing SDSS data ({\tt photo};
Lupton {\it et al.}, in preparation). {\tt photo} subtracts the CCD
bias, divides by the flat field, interpolates over data defects
(cosmic rays, bad columns, bleed trails), finds objects, deblends
overlapping objects, and performs photometry.  It estimates the local
point-spread function (PSF) as a function of position based on bright
stars. Among many other tasks, for each photometric band the pipeline
calculates the $3''$ diameter aperture magnitude $m_{\mathrm{fiber}}$,
the PSF magnitude $m_{\mathrm{PSF}}$ (that is, the magnitude using the
local PSF as a weighted aperture), and the Petrosian magnitude $m_P$
(described more fully below). It fits for the parameters of (possibly
elliptical) de Vaucouleurs and exponential profiles for each
object. It picks the better model of the two in $r^\ast$, and uses
this model as a weighted aperture to calculate the ``model magnitude''
$m_{\mathrm{model}}$ in all bands; as described in the Section
\ref{target}, while the flux limit is defined with repect to Petrosian
magnitudes, the model magnitudes are used to distinguish between stars
and galaxies. More sophicated models will be fit to a bright subsample
of the galaxies in a later paper, but the survey pipelines measure
model magnitudes for every detected object in the survey, and to fit
more complicated models to all objects would be prohibitively
expensive.  For each object, {\tt photo} also measures an azimuthally
averaged radial profile out to the maximum detectable distance.

The measured magnitudes are calibrated in the $AB_{95}$ system
(\citealt{fukugita96a}). The system is based on three ``fundamental
standards'' (BD $+ 17^\circ 4708$, BD $+ 26^\circ 2606$, and BD $+
21^\circ 609$). The United States Naval Observatory 1m telescope has
calibrated about 150 ``primary standards'' with respect to the
fundamental standards. Meanwhile, a 0.5m photometric telescope (PT;
Uomoto {\it et al.}, in preparation) is used to relate the primary
standards to a set of secondary standards, all fainter than
$14^{\mathrm{m}}$ and lying in patches referred to as ``secondary
calibration patches.''  These patches lie along the survey stripes.
Thus, during the course of its imaging on any given night, the camera
passes over these secondary calibration patches; because the secondary
standards are faint enough that they do not saturate the camera on the
2.5m, they can be used to calibrate yet fainter objects detected by
the imaging. Then, all of the detected objects can be related to the
primary standards, and thus to the fundamental standards. The PT
serves yet another function, which is to monitor the primary standards
during nights that the 2.5m telescope is imaging, in order to
determine extinction coefficients and their variation over the course
of the night. While this procedure was used to calibrate the
commissioning data, the resulting calibration has not yet been fully
verified. Here, we represent magnitudes with the notation $u^\ast$,
$g^\ast$, $r^\ast$, $i^\ast$, and $z^\ast$, rather than the standard
notation $u'$, $g'$, $r'$, $i'$, and $z'$, to indicate the preliminary
nature of the calibrations used here. We expect the final system to
differ from the current system by 0.05 magnitudes or less.  In
particular, we have found consistent results from calculating the
luminosity function in regions other than the equatorial stripe
presented here, which were calibrated based on different secondary
patches.

\subsection{Petrosian Magnitudes}
\label{petro}

Because galaxies are resolved objects with poorly defined edges, and
do not all have the same radial surface brightness profile, some care
is required to define the ``flux'' associated with each object. An
ideal method would measure the ``total'' light associated with each
galaxy, but in practice any such method requires a model-dependent
extrapolation of the measured light profile, and an accurate
extrapolation itself requires high signal-to-noise ratio data in the
outer regions of the galaxy. A conventional alternative is to measure
flux within a specified isophote, but with such isophotal magnitudes
the {\it fraction} of a galaxy's light that is measured depends on the
{\it amplitude} of its surface brightness profile, and the fraction
decreases if the surface brightness is diminished by cosmological
redshift dimming or by Galactic extinction. To avoid these problems,
the SDSS has adopted a modified form of the \citet{petrosian76a}
system, measuring galaxy fluxes with a circular aperture whose radius
is defined by the {\it shape} of the galaxy light profile.

More specifically, we define the ``Petrosian ratio'' $\petroratio$ at
a radius $r$ from the center of an object to be the ratio of the local
surface brightness averaged over an annulus at $r$ to the mean surface
brightness within $r$:
\begin{equation}
\label{petroratio}
\petroratio (r)\equiv \frac{\left.
\int_{\alpha_{\mathrm{lo}} r}^{\alpha_{\mathrm{hi}} r} dr' 2\pi r'
I(r') \right/ \left[\pi(\alpha_{\mathrm{hi}}^2 -
\alpha_{\mathrm{lo}}^2) r^2\right]}{\left.
\int_0^r dr' 2\pi r'
I(r') \right/ [\pi r^2]},
\end{equation}
where $I(r)$ is the azimuthally averaged surface brightness profile and
$\alpha_{\mathrm{lo}}<1$, $\alpha_{\mathrm{hi}}>1$ define the annulus.
The SDSS has adopted $\alpha_{\mathrm{lo}}=0.8$ and
$\alpha_{\mathrm{hi}}=1.25$.

The Petrosian radius $\petroradius$ is defined as the radius at which
$\petroratio(r_P)$ equals some specified value $\petroratiolim$. The
Petrosian flux in any band is then defined as the flux within a
certain number $\petronumber$ of $r^\ast$ Petrosian radii:
\begin{equation}
F_P \equiv \int_0^{\petronumber \petroradius} 2\pi r'dr' I(r').
\end{equation}
Thus, the aperture in all bands is set by the profile of the galaxy in
$r^\ast$ alone.  The SDSS has selected $\petroratiolim = 0.2$ and
$\petronumber = 2$.  The aperture $2r_P$ is large enough to contain
nearly all of the light for a typical galaxy profile (see below), so
even substantial errors in $r_P$ cause only small errors in the
Petrosian flux, but small enough that sky noise in $F_P$ is small
(typical statistical errors near the flux limit of $r^\ast = 17.65$ are
$< 5\%$).
In practice, there are a number of complications associated with this
definition: the galaxy images are pixelized, some galaxies have
undefined Petrosian radii because sky noise begins to dominate before
the Petrosian ratio drops to ${\mathcal{R}_P}$, some galaxies have
multiply-defined Petrosian radii because of galaxy substructure, and
so forth. We defer detailed discussion of these issues to Lupton {\it
et al.} (in preparation), since for nearly all galaxies in the 
spectroscopic sample the idealized account above is accurate.

Given the Petrosian flux, one can find the radius $r_{50}$ contained
50\% of the Petrosian flux and the radius $r_{90}$ containing 90\% of
the Petrosian flux. As shown below, the Petrosian flux measures
virtually all of the flux in pure exponential profiles but only 80\%
of the flux in pure de Vaucouleurs profiles. Thus, for exponential
profiles, $r_{50}$ and $r_{90}$ correspond to the true 50\% and 90\%
light radii, but for de Vaucouleurs profiles $r_{50,\mathrm{petro}}
\approx 0.7 r_{50,\mathrm{true}}$ and $r_{90,\mathrm{petro}} \approx
0.43 r_{90,\mathrm{true}}$. We define the half-light surface
brightness to be the average surface brightness within the half-light
radius, in magnitudes per square arcsecond: $\mu_{1/2} \equiv m_P +
2.5 \log_{10}(2 \pi r_{50}^2)$.  The ``concentration index'' of galaxies
is defined as $c\equiv r_{90}/r_{50}$. High concentration index
objects, such as de Vaucouleurs profile galaxies, have a strong
central concentration of light and large, faint wings. Low
concentration index objects, such as exponential profile galaxies,
have a light distribution which is closer to uniform. We will
generally express our results as a function of $1/c$, the inverse
concentration index.

The Petrosian ratio $\petroratio$ is manifestly indifferent to
multiplicative changes in the surface brightness of the galaxy and
depends only on the physical radius in the galaxy, not on its redshift
or Galactic extinction. Thus, the ratio of the Petrosian flux to total
flux for any given galaxy depends on redshift only to the extent that
the effects of seeing and $K$-corrections change the measured shape of
the galaxy profile. Petrosian magnitudes only become practical for
relatively deep imaging; otherwise the Petrosian ratio (Equation
\ref{petroratio}) is noisy and the aperture cannot be made large
enough to capture most of a typical galaxy's light.

\begin{figure}
\figurenum{\fignum}
\plotone{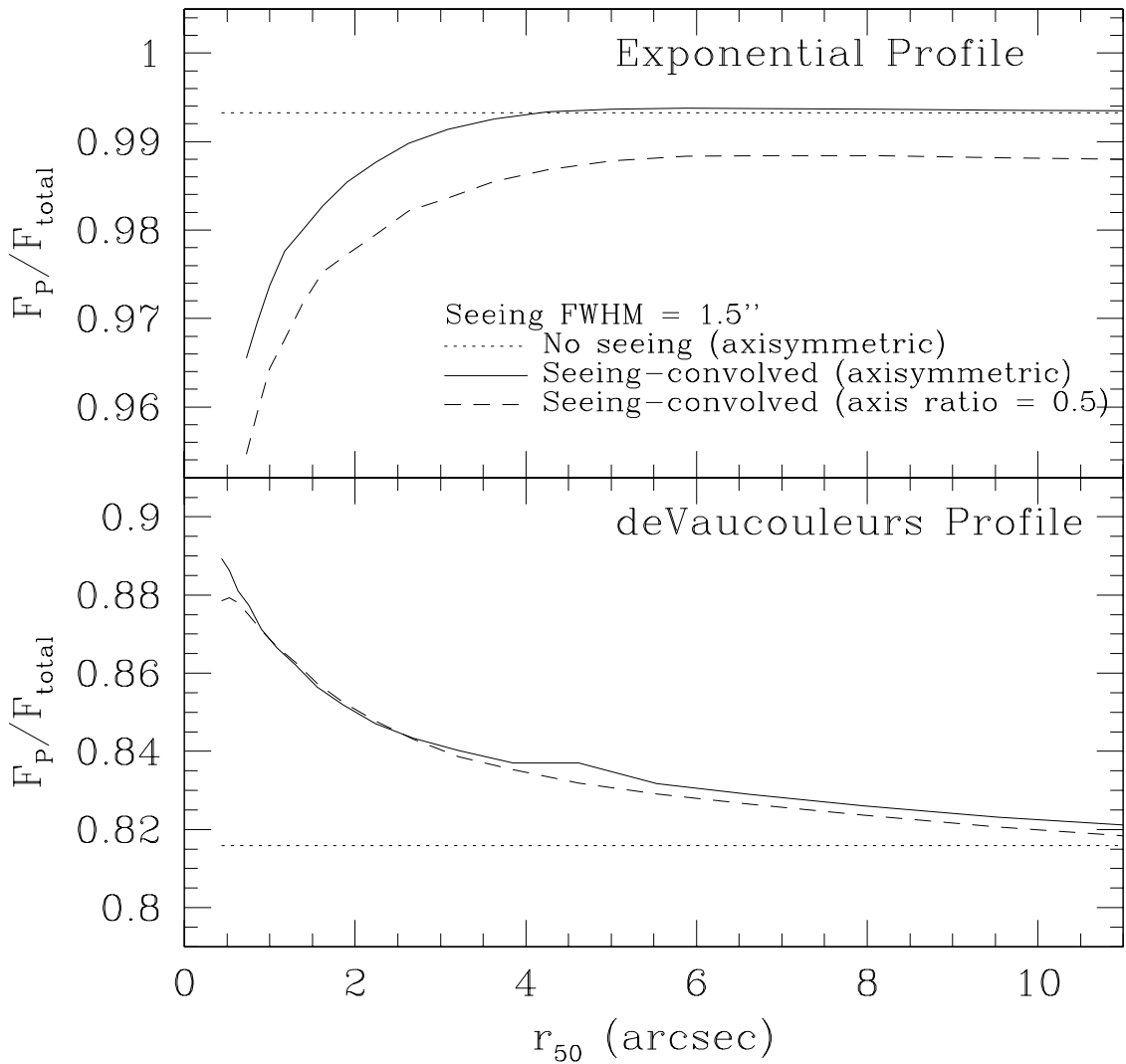}
\caption{\label{pvr.lf} Behavior of Petrosian magnitudes as a function
of galaxy size, for pure, face-on exponential profiles {\it (top
panel)} and circularly symmetric de Vaucouleurs profiles {\it (bottom
panel)}, under marginal seeing conditions in the SDSS. For each
profile, the fraction of total light measured by the Petrosian flux is
plotted against the (true) angular half-light radius of the galaxy. In
the absence of seeing, Petrosian flux measures a constant fraction of
the total light as a function of redshift, unlike isophotal
fluxes. However, small differences do appear as a function of the
angular size due to seeing. As the galaxy size becomes comparable to
the size of the seeing disk, the fraction of light measured by
Petrosian quantities approaches that fraction for a PSF, which is
about 95\%. In the case of exponential disks, this reduces the flux,
because in the absence of seeing nearly 100\% of the light is
measured. In the case of a de Vaucouleurs profile, this increases the
flux, because in the absence of seeing about 80\% of the light is
measured (assuming that the de Vaucouleurs profile extends to
infinity).  As shown in Figure \ref{r50hist}, a significant number of
galaxies have sizes close to that of the seeing disk, so this
dependence on galaxy size for small galaxies (which is present in
isophotal magnitudes as well) will limit the accuracy of our estimates
of the total luminosity density, though only to a few percent.}
\end{figure}

Using simulated galaxy observations, we can determine what fraction of
the light the Petrosian flux contains for typical galaxy
profiles. This fraction is independent of redshift except when the
size of the galaxy is comparable to the seeing.  Figure \ref{pvr.lf}
shows the ratio $F_P/F_{\mathrm{total}}$ of the Petrosian flux to the
total flux, as a function of the (true) angular half-light radius of a
galaxy. The top panel shows $F_P/F_{\mathrm{total}}$ for a face-on
exponential disk; the bottom panel shows the same quantity for a
circularly symmetric de Vaucouleurs profile. The dotted line is the
result in the limit of no seeing, for an axisymmetric galaxy. The
solid line is the result assuming the median seeing of $1.5''$ for the
runs analyzed here, again for an axisymmetric galaxy. To calculate
these quantities, we convolve each model galaxy with a realistic
point-spread function (including the power-law wings) measured from
bright stars in the survey. As the size of the galaxy becomes
comparable to the seeing, the fraction of light enclosed within a
radius of $2\petroradius$ becomes closer and closer to the
corresponding fraction for a star, which is about 95\%. In the case of
exponential profiles, $F_P\approx F_{\mathrm{total}}$ in the absence
of seeing. Seeing therefore reduces the fraction of light measured by
the Petrosian flux, though only by about 3\% even for small
galaxies. In the case of de~Vaucouleurs profiles, about 82\% of the
light is measured by the Petrosian flux in the absence of
seeing. Seeing therefore increases the fraction of light measured for
a small de~Vaucouleurs galaxy by about 10\%.  These results depend
very little on the inclination of the disk or the axis ratio of the de
Vaucouleurs profile.  For example, in Figure
\ref{pvr.lf}, the dashed line in each panel shows the result for a
galaxy with an axis ratio of 0.5, convolved with seeing. 

\begin{figure}
\figurenum{\fignum}
\plotone{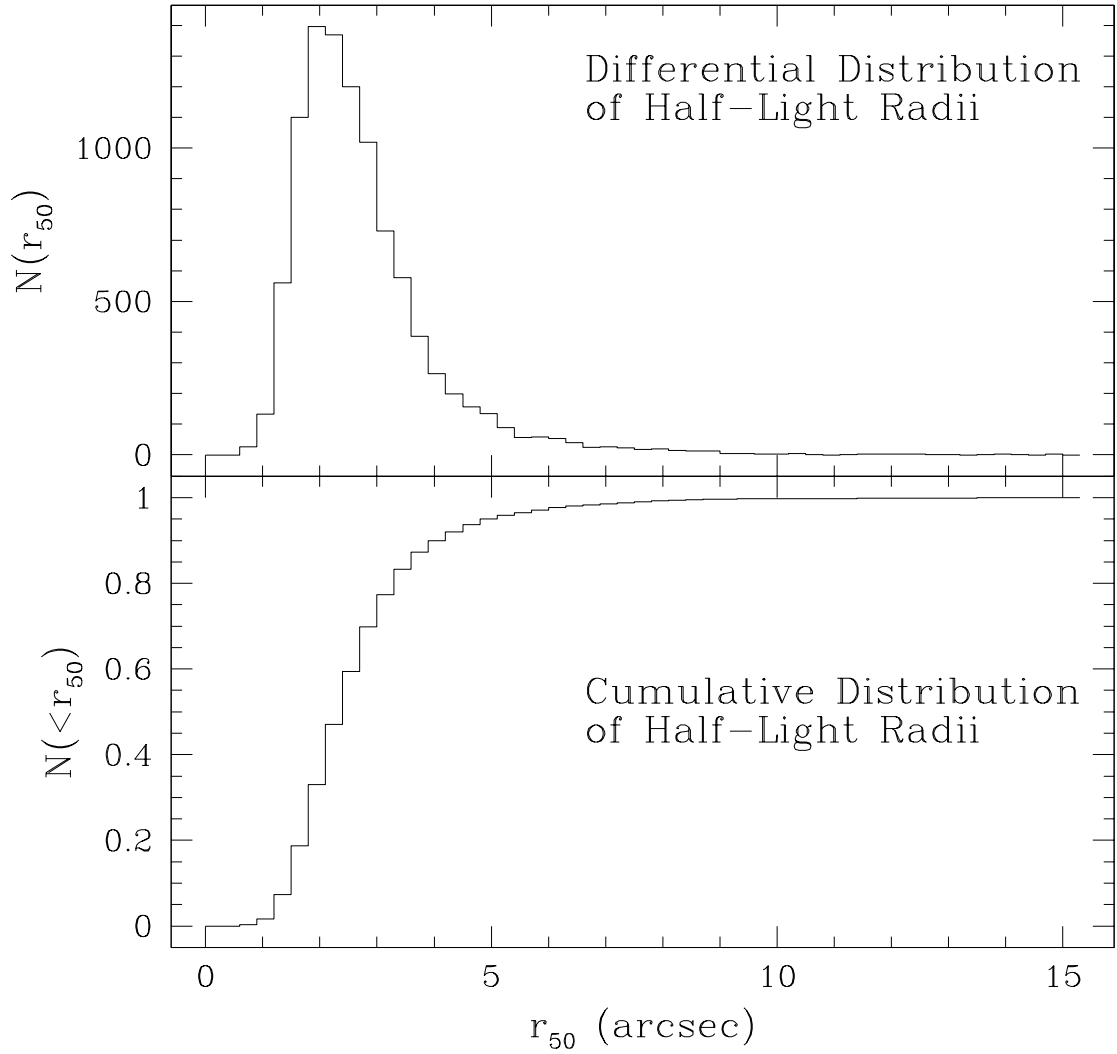}
\caption{\label{r50hist} Distribution of half-light radii $r_{50}$ of
SDSS galaxies with $r^\ast < 17.6$. Top panel shows the distribution;
bottom panel shows the fractional cumulative distribution.  Roughly
half of the galaxies are in the regime ($r_{50}<2.5''$) in which seeing
significantly affects the determination of the Petrosian radius and
flux.}
\end{figure}

For comparison, we show in Figure \ref{r50hist} the distribution of
observed half-light radii. This figure shows the distribution of
$r_{50}$ for galaxies with $r^\ast < 17.6$ in the top panel, and the
fractional cumulative distribution $N(<r_{50})$ in the bottom
panel. Roughly half of the galaxies have $r_{50}<2.5''$, and thus
probably have their Petrosian fluxes under- or overestimated by a few
percent.  We emphasize that a comparably deep isophotal magnitude
would have a similar dependence on angular size to that found here,
without the benefits of being independent of galaxy redshift for
galaxies with half-light radii larger than the seeing.

\subsection{Targeting Galaxies} 
\label{target}

\begin{figure}
\figurenum{\fignum}
\plotone{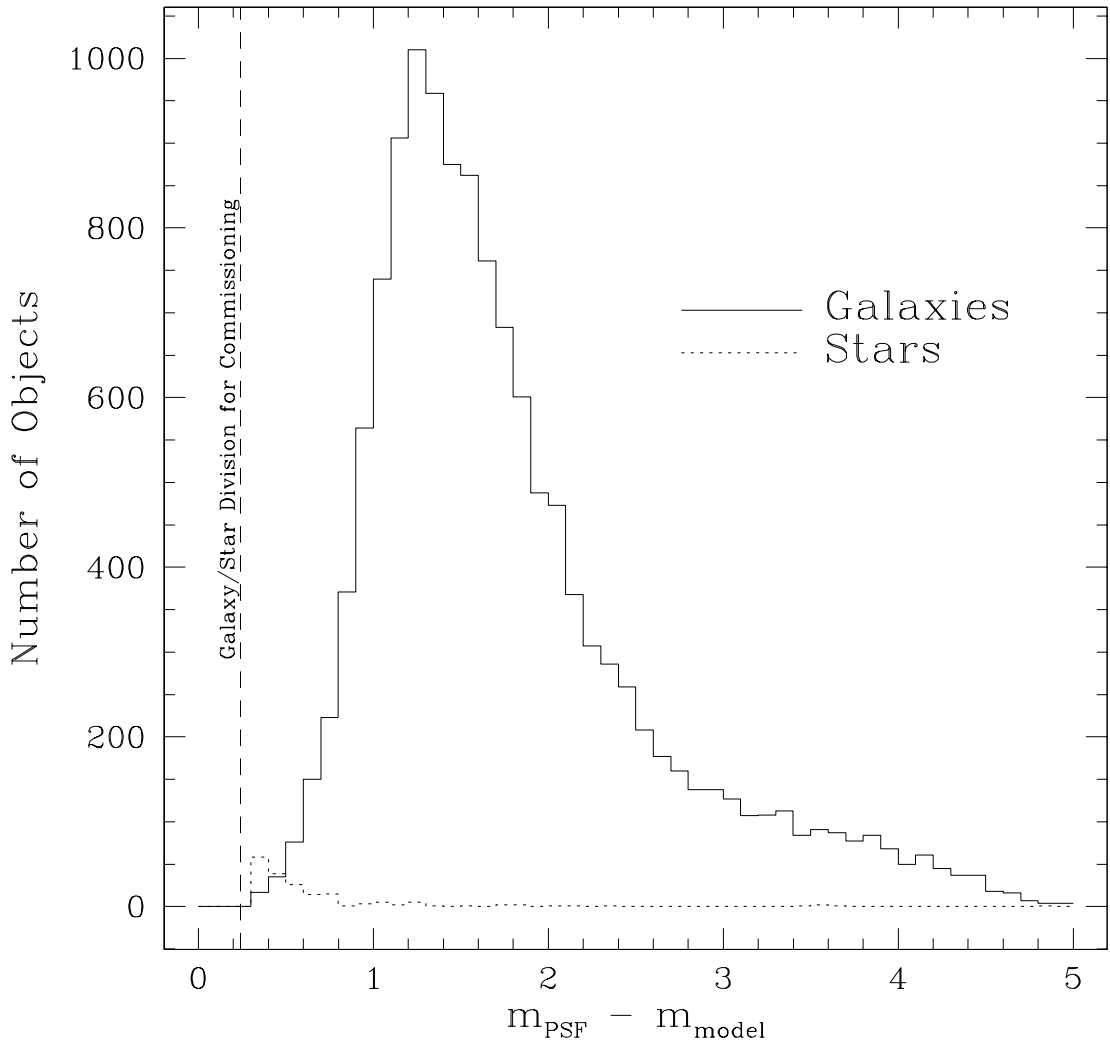}
\caption{\label{psfmodel} Distribution of the quantity
$m_{\mathrm{PSF}}-m_{\mathrm{model}}$ (determined in $r^\ast$), which
is used to separate stars from galaxies for the purposes of target
selection in the SDSS. The cut-off of 0.242 used in the commissioning
data for this purpose is shown as the vertical line. The solid
histogram is the distribution of $m_{\mathrm{PSF}}-m_{\mathrm{model}}$
for targeted objects which were spectroscopically identified as
galaxies; the dotted histogram is the distribution for targeted
objects which were spectroscopically identified as stars. About 1\% of
objects targeted as galaxies turned out to be stars. The distribution
falls off sharply at small $m_{\mathrm{PSF}}-m_{\mathrm{model}}$,
implying that $\ll 1\%$ of galaxies brighter than $r^\ast=17.6$ are
misclassified as stars. }
\end{figure}

Once objects have been identified and photometry has been performed,
which can be done reliably to $r^\ast\sim 23$, a sample of targets is
selected for spectroscopic observation. The details of this selection
and the software pipeline used to implement it will be described in
separate papers (Strauss {\it et al.}, in preparation; vanden Berk
{\it et al.}, in preparation). For the purposes of the main galaxy
sample, extended objects (galaxies) are separated from stars based on
the quantity $m_{\mathrm{PSF}}-m_{\mathrm{model}}$ in $r^\ast$, where
(as described in Section
\ref{photo}) $m_{\mathrm{PSF}}$ is the magnitude using a PSF-weighted
aperture and $m_{\mathrm{model}}$ is the magnitude associated with the
best-fit model to the galaxy profile. That is, if most of the object's
light is contained within an aperture weighted by the PSF, it is
considered to be stellar.  
Figure \ref{psfmodel} shows the
distribution of our spectroscopic galaxy targets versus
$m_{\mathrm{PSF}}-m_{\mathrm{model}}$. The objects which are galaxies
(based on the spectroscopic data) are shown in the solid histogram,
and those which are stars are shown in the dotted histogram.  For the
commissioning data, the division was set at $0.242$; roughly 1\% of
objects brighter than the spectroscopic limit and identified as
galaxies in this way are stars. Indeed, most of the stellar spectra
recovered from the galaxy sample are actually double stars, which
shows that this method of separating the populations is fairly clean.
It is clear from the drop-off in the distribution of galaxies as
$m_{\mathrm{PSF}}-m_{\mathrm{model}}$ decreases that galaxies in this
magnitude range with $m_{\mathrm{PSF}}-m_{\mathrm{model}}<0.242$ are
extremely rare.
We use the model magnitude here for
star/galaxy separation rather than the Petrosian magnitude because
the model magnitude is more robust at faint magnitudes ($r^\ast>21$)
and to the effects of seeing; we emphasize that for the spectroscopic
sample such considerations are irrelevant, because the star/galaxy
separation is so clean.

Given the set of extended objects in the commissioning data, we chose
as spectroscopic targets all objects with a Petrosian magnitude (after
an extinction correction using the dust maps of \citealt{schlegel98a})
of $r^\ast<17.65$ and a fraction of those between 17.65 and 17.75,
with a probability declining linearly from unity to zero in this
range. This flux limit was applied using the calibration available in
the March 2000, in a version of the processing known as ``rerun 1.''
The magnitudes quoted here were calibrated in a later version known as
``rerun 4.'' Because of the differences in calibration, to define a
complete sample we have restricted ourselves here to a flux limit of
$r^\ast<17.6$.

In addition, a surface brightness cut of $\mu_{1/2} < 23.5$ was used
for the commissioning observations; this cut eliminates less than 1\%
of all extended objects brighter than the flux limit, about half of
which are real galaxies and half of which are ghosts or consequences
of improper deblending. We have checked the effects of improper
deblending by examining by eye all of the targets in the range $22.5 <
\mu_{1/2} < 23.5$. Fifteen (about 3\%) of the targets in this range
are products of deblending failures which nevertheless have galactic
redshifts successfully determined and thus contaminate our sample; we
have checked that the effect of this contamination is small compared
to our estimated errors.  The main survey uses improved methods for
eliminating contamination by such defects and will accordingly have a
somewhat more relaxed surface brightness limit. We will explore below
(Section \ref{sbdep}) some of the consequences of the surface
brightness limit in the commissioning data.

A number of other requirements were used in selecting spectroscopic
targets. The ``fiber magnitude'' (defined in Section \ref{photo} as
the magnitude within an $3''$ diameter aperture) was restricted to be
fainter than $15$ in $g^\ast$, $r^\ast$, or $i^\ast$ to avoid
cross-talk between fibers on the spectrograph CCDs.
Objects brighter than $r^\ast=15.5$ and with half-light radii $< 2''$
almost invariably turn out to be misclassified stars and therefore we
reject them; this cut always affects $<1\%$ of objects, and only
approaches this percentage in poor seeing conditions.  We note that
the BRG sample targets galaxies fainter than $r^\ast =17.65$ that
satisfy specified color and magnitude selection criteria; we have {\it
not} made use of the BRG sample in the luminosity function
measurements presented here.


\subsection{Spectroscopic Observations}

The spectra of the chosen galaxies are obtained using a multi-object
fiber spectrograph that can take 640 spectra at once in a circular
plate (or ``tile'') 1.49 degrees in radius (Uomoto {\it et al.}, in
preparation).  Each exposure is about 45---60 minutes, depending on
observing conditions.  Each fiber has a diameter of $3''$, and is led
into a dichroic beam-splitter centered at $\approx 6000$ \AA\ which
separates the light onto red and blue cameras.  The resulting spectrum
covers the range from 3800 \AA\ to 9200 \AA, with a resolution $R\sim
2000$.

In practice, 48 of the fibers are used as sky fibers and
spectrophotometric standards, leaving 592 for quasars, galaxies, and
stars. Typically, about 500 fibers per plate are used for
galaxies. Although this number provides enough fibers for all our
desired targets, the galaxies are distributed non-uniformly on the
sky. Furthermore, no two fibers on the same plate can be placed closer
than $55''$, a limit imposed by the physical size of the fiber
plugs. For these two reasons, we must position the centers of the
spectroscopic plates and allocate the targets to the plates carefully,
in order to maximize the number of targeted galaxies (Blanton {\it et
al.}, in preparation). In the commissioning data we assigned fibers to
about $92\%$ of the targeted objects (lumping main sample galaxies,
QSOs, and BRGs together), although for the survey proper, for which
the geometry will be considerably more favorable, our sampling will be
above $95\%$. Most of the missing targets are due to the $55''$
collision constraints.


After observation, the spectra are analyzed using a specially written
spectroscopic pipeline (Frieman {\it et al.}, in preparation) which
extracts the 640 spectra from each two-dimensional spectrogram,
carries out wavelength and flux calibration, subtracts the sky,
classifies the objects as stars, QSOs, or galaxies, and determines the
redshift. For galaxies, redshifts are determined independently from
emission lines and from a cross-correlation with absorption spectra;
using both methods provides a useful check on the software. In the
data analyzed here (not all of which is considered ``survey
quality''), and with a commissioning version of the pipeline, the
redshift success rate for objects targeted as galaxies is about 98\%,
as determined by eyeball examination of the spectra, independent of
apparent magnitude and surface brightness. The remaining 2\% fail to
give redshifts, due for the most part to broken fibers and the
accidental targetting of ghost images. The redshift errors are around
$30$ km s$^{-1}$, as determined by repeat observations of the same
objects; this error is completely negligible for the purposes of
measuring the luminosity function.

\subsection{Sampling Rate}
\label{sampling}

To calculate the luminosity function, we must map the selection
function of the survey. Aside from the flux limit, there are three
main effects: missing galaxies due to lack of fibers in dense regions,
missing galaxies due to spectroscopic failures, and missing galaxies
due to fiber collisions.  To account for these effects, we calculate
the local sampling rate of galaxies separately for each region covered
by a unique set of plates. That is, in the case of two overlapping
plates, the sampling is calculated separately in three regions: the
region covered only by the first plate, the region covered only by the
second plate, and the region covered by both. We refer to each of
these three regions as an ``overlap region.'' Thus, depending on what
overlap region it was in, each galaxy can be assigned a sampling rate
$\tilde f_t$, equal to the number of redshifts of galaxy targets
obtained in the region divided by the number of galaxy targets in the
region. In regions covered by a single plate, $\tilde f_t \sim
0.85$--$0.9$; in multiple plate regions, $\tilde f_t \sim
0.95$--$1$. These completenesses average to the 92\% quoted above.

It is possible to simply weight each galaxy by $1/\tilde f_t$ when
calculating the luminosity function. However, the majority of the
missing galaxies are missing because of the fiber collision
constraints. It is more appropriate to account for the collisions as
follows. First, we group the galaxies using a friends-of-friends
algorithm with a $55''$ linking length. For each galaxy for which we
obtained a redshift, we assign a ``collision weight'' $w_c$ equal to
the multiplicity of the group divided by the number of redshifts
obtained in the group. That is, in a colliding pair for which we have
one redshift, we weight that redshift by a factor $w_c=2$.  This
procedure is essentially the same as assuming the two galaxies have
the same absolute magnitude.  The sampling rate using these collision
weights, which we denote $f_t$, turns out to be $\sim 0.97$--$1$ for
most overlap regions. When we calculate the normalization of the
luminosity function below, we weight each galaxy by the factor
$w_c/f_t$. We should note that the difference between weighting by
$w_c/f_t$ and by $1/\tilde f_t$ is negligible in the calculation of
the luminosity function (although this is certainly not the case for
clustering statistics).

\subsection{Distance Modulus and $K$-corrections}

To convert the apparent magnitude $m$ to an absolute magnitude $M$, we
must assume a particular cosmology and correct for the fact that the
observed bandpass differs from the restframe bandpass. The relation
between $m$ and $M$ can be written
\begin{equation}
M = m - \mathrm{DM}(z) - K(z), 
\end{equation}
where $\mathrm{DM}(z)$ is the bolometric distance modulus and $K(z)$
is the $K$-correction.

Here we use distance moduli based on three different
Friedman-Robertson-Walker models: ($\Omega_m=1$, $\Omega_\Lambda =
0$); ($\Omega_m=0.3$, $\Omega_\Lambda = 0.7$); and ($\Omega_m=0.3$,
$\Omega_\Lambda = 0$). See \citet{hogg99a} and references therein for
distance measure formulae for these cosmologies.  At $z=0.2$, the
highest redshift considered here, the difference in the distance
modulus for these three cases is about 0.2 magnitudes at
most. Meanwhile, the comoving volume associated with a coordinate
volume element $d\Omega dz$ differs among these models by as much as
40\%. We use all three models for the calculation of our main results
for the luminosity in each band, but for the investigation of the
dependence of luminosity on other parameters, we consider only the
$\Omega_m=0.3$, $\Omega_\Lambda = 0.7$ model for the sake of
simplicity. We use $H_0 = 100 h$ km s$^{-1}$ Mpc$^{-1}$, with $h=1$
when we have not specified $h$ explicitly; choosing alternative values
for $h$ shifts all absolute magnitudes by $5\log_{10}h$ and changes
number densities by the factor $h^3$.

\begin{figure}
\figurenum{\fignum}
\plotone{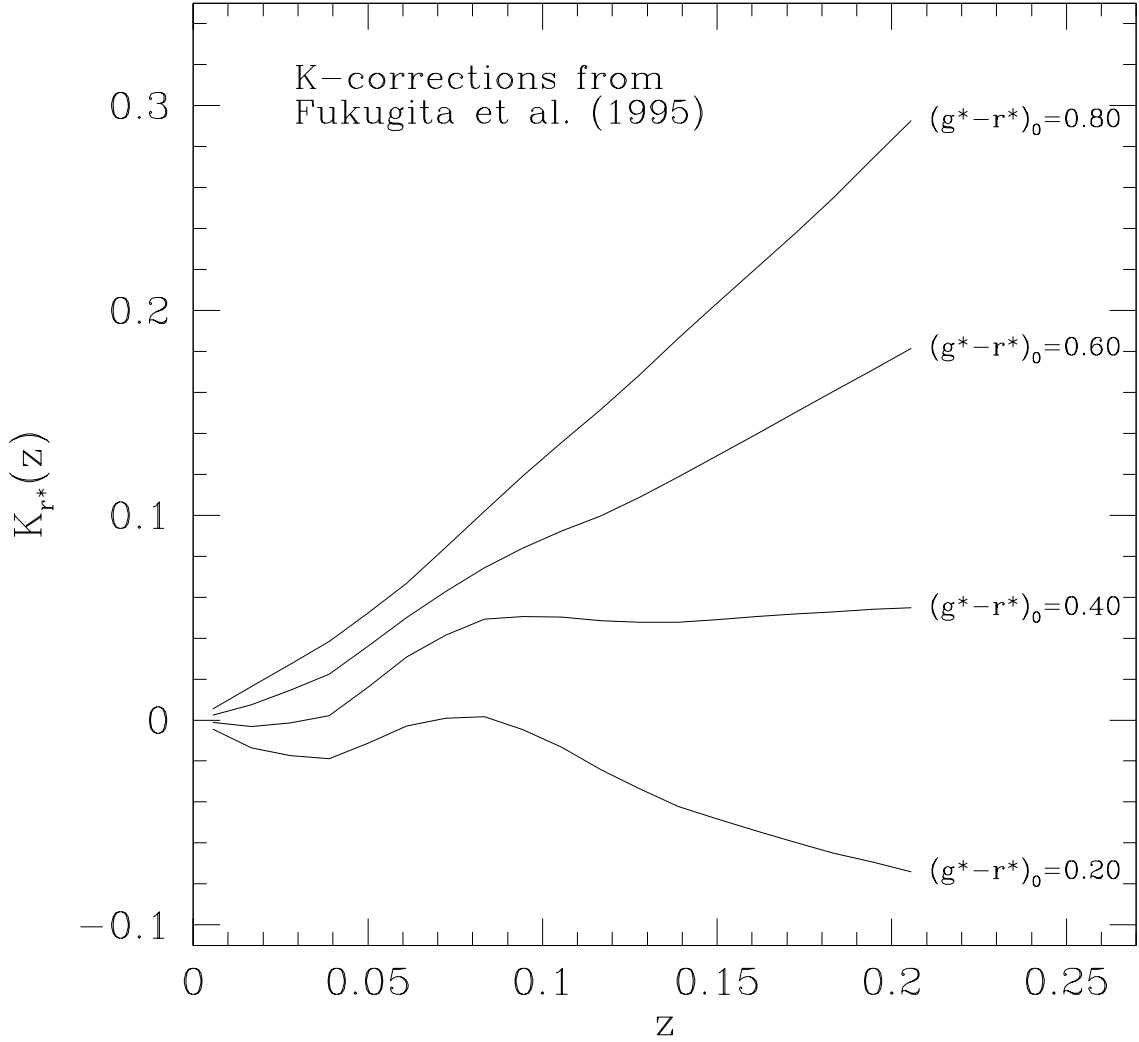}
\caption{\label{kcorrect} $K$-corrections in the $r^\ast$ band for
galaxies with intrinsic $g^\ast-r^\ast$ colors of 0.2, 0.4, 0.6, and
0.8, as labeled. These curves are obtained by linear interpolation
between the morphological types used by \citet{fukugita95a}. Similar
results are available for the other four bands.  }
\end{figure}

Because we observe the galaxies in a fixed observed wavelength band, a
$K$-correction is necessary to account for the fact that the observed
band corresponds to different rest-frame bands at different redshifts.
Naturally, this $K$-correction is dependent on the galaxy spectral
energy distribution (SED). Given the five-band SDSS photometry and the
high quality spectra over a large range in wavelength, the best
approach in the long run will be to determine the $K$-corrections of
the galaxies from the survey itself.  For the present, however, we
will simply use $K$-corrections supplied by \citet{fukugita95a} for a
range of galaxy types.  Using the observed galaxy redshift and the
$g^\ast-r^\ast$ color, we interpolate between the galaxy types in
\citet{fukugita95a} to determine the intrinsic $g^\ast-r^\ast$ color
and the appropriate $K$-correction.  Figure \ref{kcorrect} shows the
$K$-correction in $r^\ast$ as a function of redshift for several
different galaxy colors over the range of redshifts of interest here.
While the true survey bandpasses are not exactly those assumed by
\citet{fukugita95a}, we have tested the effect of using
$K$-corrections based on recent estimates of the filter curves
(\citealt{fan01a}), and the differences in the resulting measured
luminosity density are less than 5\% in all bands.

\section{Calculating the Luminosity Function}
\label{method}

We define the unit-normalized luminosity function $\Phi(L)$ to be the
distribution function of galaxies in luminosity, normalized to unit
integral over the range from $L_{\mathrm{min}}$ to
$L_{\mathrm{max}}$. We denote the number density of galaxies in the
range of luminosity considered as ${\bar n}$, in units of $h^3$
Mpc$^{-3}$. Thus, the number density of galaxies per unit luminosity
is $\hat\Phi(L) = {\bar n} \Phi(L)$. For plotting purposes, we use the
luminosity function per unit magnitude, normalized to the mean density
of galaxies:
\begin{equation}
\hat\Phi(M) = 0.4 \ln(10) {\bar n}  L \Phi(L),
\end{equation}
with $M=-2.5 \log_{10} (L/L_0)$, where $L_0$ is the luminosity of an
object with absolute magnitude zero.

We use the maximum likelihood approach outlined by \citet{efstathiou88a}
and \citet{sandage79a} to calculate $\Phi(L)$.  This method is based on
the conditional probability of observing a galaxy with a luminosity
between $L_j$ and $L_j+dL_j$ given its redshift $z_j$:
\begin{equation}
\label{likefunc}
p(L_j| z_j) dL_j = \frac{p(L_j,z_j) dL_j dz_j}{p(z_j) dz_j} = 
\frac{\Phi(L_j) }
{\int_{L_{{\mathrm{min}}}(z_j)}^{L_{{\mathrm{max}}}(z_j)} dL \Phi(L) }
dL_j.
\end{equation}
Here, $L_{\mathrm{min}}(z_j)$ and $L_{\mathrm{max}}(z_j)$ are the
minimum and maximum luminosities observable at redshift $z_j$, given
the faint and bright flux limits of the survey. As described below, we
evaluate Equation (\ref{likefunc}) using two models for $\Phi(L)$: a
Schechter function and a non-parametric model.  Note that the
probability in Equation (\ref{likefunc}) is completely independent of
the density field of galaxies; this fact makes the method insensitive
to the effects of large-scale structure, at least to the extent that
the luminosity function is universal. In addition, this insensitivity
makes it necessary to calculate the normalization in a separate step,
described below.

The procedure is to minimize the quantity
\begin{equation}
{\mathcal L} = - 2 \sum_j w_c \log\left[p(L_j | z_j)\right],
\end{equation}
where the weight $w_c$ is applied to properly account for galaxies
which are eliminated due to fiber collisions. That is, we assign
double weight to a galaxy if it has eliminated a neighbor from the
redshift sample due to a collision. In principle, we should use the
weight $w_c/f_t$, to also weight those galaxies that do not have
fibers due to being in dense regions, but the changes in our results
would be extremely small. Note that the weights we assign only affect
the mean of the maximum likelihood estimator if they correlate with
the luminosity; that is, the weights $w_c$ only make a difference to
the extent that the typical luminosities of galaxies involved in fiber
collisions are different from those of other galaxies.

To perform the minimization, we need to choose a model for
$\Phi(L)$. Here we use both a \citet{schechter76a} function and a
non-parametric model. The Schechter function is 
\begin{equation}
\label{schechterL}
\hat\Phi(L) dL = \phi_\ast 
\left(\frac{L}{L_\ast}\right)^{\alpha}
\exp\left(-L/L_\ast\right) dL.
\end{equation}
Expressed per unit magnitude, it is
\begin{equation}
\label{schechter}
\hat\Phi(M) dM = 
0.4 \ln(10)
\phi_\ast 
10^{-0.4(M-M_\ast)(\alpha+1)} \exp\left[- 10^{-0.4(M-M_\ast)}\right]
dM.
\end{equation}
In this case, the parameter space of $M_\ast$ and $\alpha$ is searched
to minimize ${\mathcal L}$.  To calculate the error bars in this case,
we perform 200 Monte Carlo calculations of the following form. We take
all the galaxy positions as observed in the survey, and for each assign
a random luminosity drawn from the fitted luminosity function and the
flux limits. Then we fit for the best Schechter function in each Monte
Carlo sample and use the variation among samples to calculate the errors
and covariance of $M_\ast$ and $\alpha$.

We have checked the difference between the Monte Carlo error estimate
and a jackknife error estimate (\citealt{lupton93a}). For the jackknife
method, we divide our sample into eighteen regions on the sky of
approximately equal area. Then we perform the luminosity function
analysis eighteen times, each time leaving a different section
out. The estimated statistical variance of a parameter $x$ is
\begin{equation}
\mathrm{Var}(x) = \frac{N-1}{N} \sum_{i=1}^{N} (x - \bar x)^2,
\end{equation}
where $N=18$ in this case and $\bar x$ is the mean value of the
parameter measured in the samples. The errors derived in this way are
within 30\% of those found using the Monte Carlo method described in the
previous paragraph.

The second model for $\Phi(L)$ we use is the non-parametric model,
which is a piece-wise constant interpolation in logarithmically spaced
steps in luminosity, as described by
\citet{efstathiou88a}. For this model, the likelihood given by Equation
(\ref{likefunc}) can be maximized quickly by using an iterative
equation.
In this case, because there are so many parameters to constrain in the
covariance matrix, we follow \citet{efstathiou88a} and calculate the
errors based on the second derivatives of the likelihood function
around the solution.

Given the fit to either model, we need to calculate the
normalization. We do so here using the minimum variance estimator of
\citet{davis82a}:
\begin{equation}
\label{normalization}
{\bar n} = \frac{\sum_{j=1}^{N_{\mathrm{gals}}} w_c w(z_j)}
{\int dV \phi(z) w(z)},
\end{equation}
where the integral is over the volume covered by the survey between the
minimum and maximum redshifts used for our estimate.  The weight for
each galaxy is
\begin{equation}
w(z)=\frac{1}{1+ {\bar n} J_3 \phi(z)},
\end{equation}
and the selection function is
\begin{equation}
\label{selfunc}
\phi(z) = {\int_{L_{\mathrm{min}}(z)}^{L_{\mathrm{max}}(z)} dL\, \Phi(L)
f_{t}},
\end{equation}
where $f_t$ is the galaxy sampling rate as described above. The
integral of the correlation function is:
\begin{equation}
J_3 = \int_0^\infty dr\, r^2 \xi(r).
\end{equation}
Clearly, because ${\bar n}$ appears in the weight $w(z)$, it must be
determined iteratively, which we do using the simple estimator ${\bar n}
= (1/V) \sum w_c/\phi(z_j)$ as an initial guess. We iterate until the
change is less than 10\% of the estimated error. The error in the
normalization is given by
\begin{equation}
\label{normerr}
\Avg{\delta {\bar n}^2}^{1/2} = \left[
\frac{{\bar n}}{\int dV \phi(z) w(z)}
\right]^{1/2}.
\end{equation}
We have also checked the error estimates in the normalization using the
jackknife method described above and find the results to be consistent
with those of Equation (\ref{normerr}).  In addition, there are hints in
the jackknife results that the errors in ${\bar n}$ may be correlated
with the errors in $M_\ast$ and $\alpha$. However, eighteen jackknife
samples still provide a rather noisy estimate of the six parameters in a
$3\times 3$ covariance matrix. We leave the question of these correlated
errors for a future analysis of a larger data set.

The luminosities of galaxies are known to correlate with other
quantities, such as their surface brightness, color, and morphology.  We
investigate these relationships below and will need a way to calculate
the joint relationship between luminosity and other quantities. To do
so,
we use a two-dimensional generalization of the method of
\citet{efstathiou88a}, along the lines of \citet{sodre93a},
\citet{dejong00a}, and \citet{oneil99a} (all of whom considered the
specific case of galaxy surface brightnesses). Instead of bins in
luminosity, we take bins in the two-dimensional plane of luminosity and
the other quantity in question. The derivation of the method of
\citet{efstathiou88a} and its implementation are virtually unchanged
from the one-dimensional case.

The redshift and flux limits we apply in each band are tabulated in
Table \ref{lf_limits}.  The high redshift limit is imposed to minimize
noise in the determination of the normalization.  The low redshift
limit is applied to avoid difficulties in the commissioning data
associated with the accidental deblending of nearby, large
galaxies. While the flux limit in $r^\ast$ is set by the spectroscopic
flux limit, we calculate the luminosity function in other bands by
defining flux limits in these bands sufficiently bright that almost
all of the galaxies brighter than the limit were actually
targeted. Imposing the requirement that $<2\%$ of galaxies brighter
than the flux limit in the given band are fainter than $r^\ast=17.6$
yields the flux limits of $u^\ast_{\mathrm{lim}} = 18.4$,
$g^\ast_{\mathrm{lim}}=17.65$, $i^\ast_{\mathrm{lim}}=16.9$, and
$z^\ast_{\mathrm{lim}}=16.5$. That is, for $u^\ast$ and $g^\ast$ we
make sure that the flux limit is bright enough to include all but the
very bluest galaxies at $r^\ast=17.6$, and for $i^\ast$ and $z^\ast$
we make sure it is bright enough to include all but the very reddest
galaxies. This approach guarantees that the full range of galaxy
colors is represented at each apparent magnitude in each sample.
Neglecting the dependence of the $K$-corrections on galaxy color, this
would be the equivalent of considering nearly the full range of galaxy
colors at each redshift.  Note that for $u^\ast$ and $g^\ast$, the
flux limits are shallow enough that we need to reduce the maximum
redshift we consider. To use a larger fraction of the data to obtain
the same result, we could calculate the bivariate luminosity function
in $r^\ast$ and in the other band of interest, but for the moment we
will stick to the simple and conservative approach.

\section{Results for the Luminosity Function}
\label{results}

\subsection{Luminosity Function in Five Bands}
\label{lffunc}

\begin{figure}
\figurenum{\fignum}
\plotone{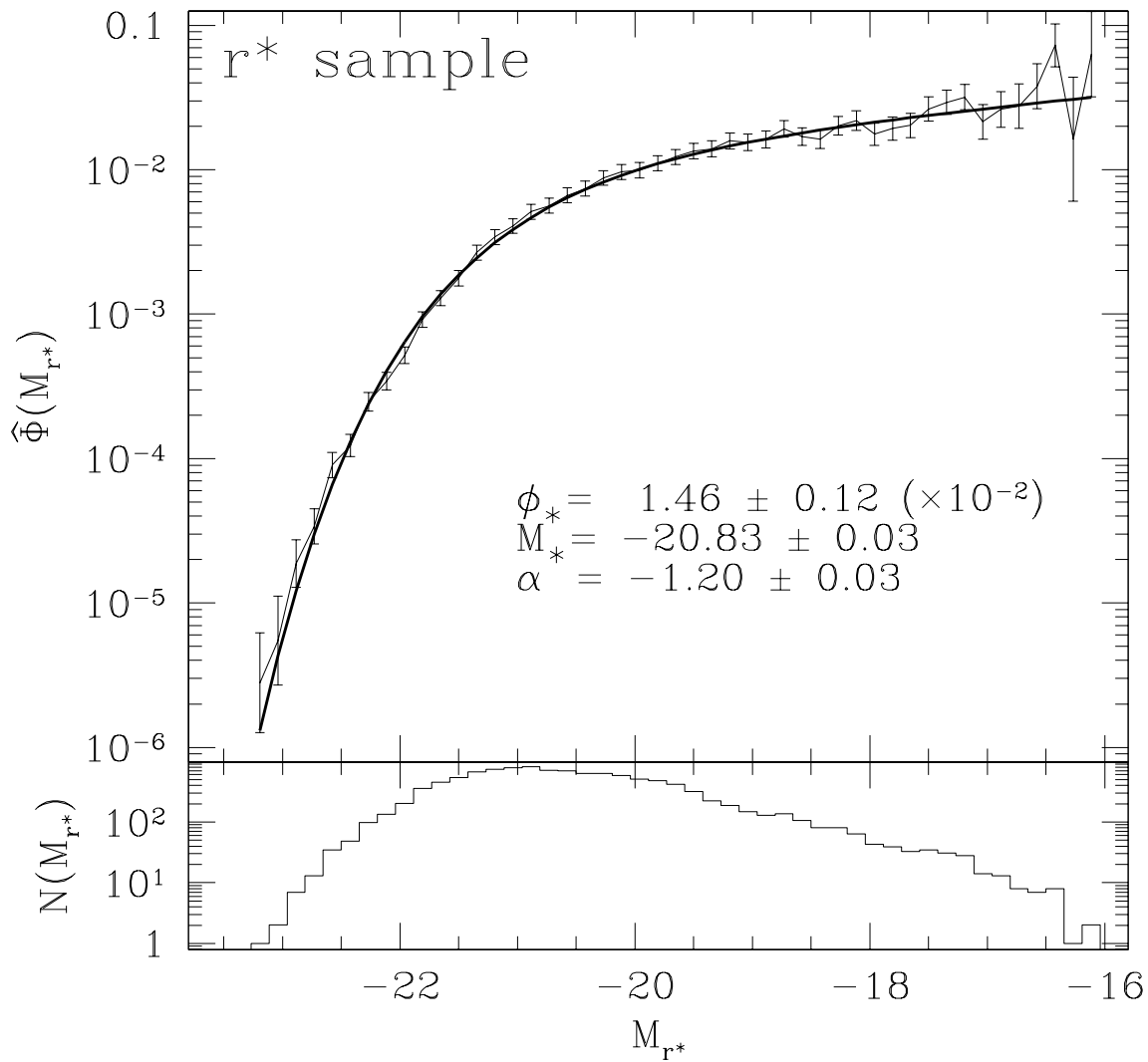}
\caption{\label{lf_full_r} Top panel shows the luminosity function of 
SDSS galaxies in the $r^\ast$ band, based on a sample of $11,275$
galaxies in the commissioning data. For these results, we assumed
$\Omega_m=0.3$ and $\Omega_\Lambda=0.7$. Results for other cosmologies
are given in Table
\ref{lf_table}. The Schechter fit ({\it thick line}) and the
non-parametric fit ({\it thin line with error bars}) agree well. The
correlation coefficient between $\alpha$ and $M_\ast$ is $r=0.78$,
showing that the errors in these parameters are strongly
correlated. Bottom panel shows the number of galaxies in each bin used
to determine the luminosity function; even though the least luminous
galaxies have the highest space density, there are very few such
galaxies in a flux-limited sample. The fact that the low-luminosity
end of the correlation function is based on a relatively small
fraction of the galaxies in the sample is the reason that the errors
in the normalization of the luminosity function are not comparable to
the Poisson errors expected given the number of galaxies; that is, a
small number of the galaxies are given disproportionate statistical
weight, making the errors greater than Poisson.}
\end{figure}

Using the galaxy sample described in Section \ref{data}, we fit the
models for the luminosity function described in Section \ref{method}.
The results for the $r^\ast$ band are shown in Figure \ref{lf_full_r}
for the case of the $\Omega_m=0.3$, $\Omega_\Lambda=0.7$
cosmology. The parameters of the fit to the Schechter function for all
three models are given in Table \ref{lf_table}, along with their
errors and covariances. The covariance is expressed as the correlation
coefficient $r$ between the errors. As previous authors have found,
the errors in $M_\ast$ and $\alpha$ have a strong correlation. Judging
from the non-parametric results, the Schechter function is a good fit
to the data. In fact, from the likelihoods calculated in the fits, we
found that the Schechter function is a {\it better} fit than the
non-parametric fit. This result is almost certainly due to the fact
that the non-parametric method assumes that the luminosity function is
flat within each bin; for example, if we double the number of
luminosity bins for the non-parametric fit, then its likelihood is
higher than that of the Schechter function fit.  Future work will
investigate this issue further using non-parametric methods like those
of \citet{koranyi97a} and \citet{springel98a}, which adopt more
general interpolation schemes than the standard piece-wise constant
interpolation of \citet{efstathiou88a}.

\begin{figure}
\figurenum{\fignum}
\plotone{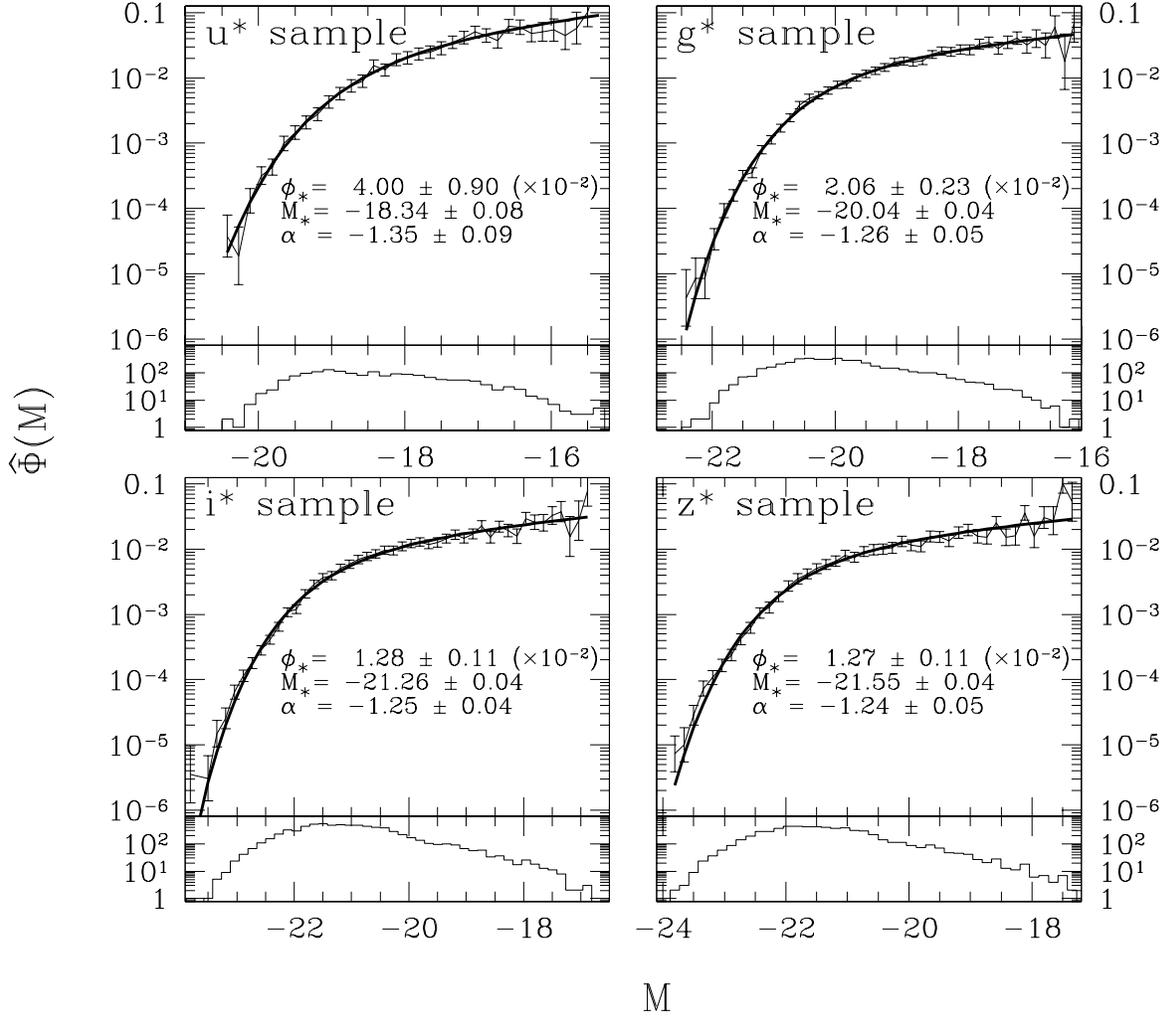}
\caption{\label{lf_full_ugiz} Same as Figure \ref{lf_full_r}, for
the $u^\ast$, $g^\ast$, $i^\ast$, and $z^\ast$ bands. The flux limits
and number of galaxies in each sample are listed in Table
\ref{lf_table}.  As in the $r^\ast$ luminosity function in Figure
\ref{lf_full_r}, the Schechter fits ({\it thick line}) and the
non-parametric fits ({\it thin line with error bars}) agree well. }
\end{figure}

We show the results of our non-parametric fits and Schechter function
fits for each of the four other bands in Figure \ref{lf_full_ugiz} and
Table \ref{lf_table}.  The Schechter function fit is good in every
band; its slope $\alpha$ for low luminosity galaxies is somewhat
steeper in $u^\ast$, but in general it is a remarkably weak function
of bandpass.  The non-parametric fits are well-constrained down to
absolute magnitudes of $-16$ to $-17.5$, depending on bandpass, about
four magnitudes less luminous than $M_\ast$ in the central ($g^\ast$,
$r^\ast$, $i^\ast$) bands.

If we extrapolate the fitted Schechter function to low luminosities,
the luminosity density can be calculated:
\begin{eqnarray}
j &=& \int_0^\infty dL L \Phi(L)\cr
&=& \phi_\ast L_\ast \Gamma(\alpha+2).
\end{eqnarray}
Similarly, one can add up the luminosity in the non-parametric fit
over the absolute magnitude range that we have measured it. We list in
Table \ref{lf_table} the luminosity density $j$ found for the
extrapolated Schechter function in each band (in units of absolute
magnitude per $h^{-3}$ Mpc$^3$) and the fraction of this luminosity
$f_{\mathrm{np}}$ that is accounted for in the non-parametric fit.
The errors given in $j$ are calculated from the errors in $\alpha$, 
$M_\ast$, and $\phi_\ast$, accounting properly for the covariance
between the errors of $\alpha$ and $M_\ast$ but assuming the errors in
$\phi_\ast$ are independent; these errors are consistent with the
errors estimated based on the jackknife samples described in Section
\ref{method}. From these results, we see that if the 
fitted slope $\alpha$ continues
to describe the luminosity function at low luminosities, the galaxies
represented in our sample account for most of the luminosity density
of the universe. We will discuss the caveats associated with surface
brightness selection effects later in Section \ref{sbdep}.

Table \ref{lf_table} also lists the luminosity density expressed in
solar units. To determine these quantities, we assume
$M_{V\odot}=4.83$, $(B-V)_{\odot}=0.65$, $(U-B)_{\odot}=0.13$, and
$(R-I)_{\odot}=0.34$ (\citealt{binney98a}), and use the stellar color
transformations of
\citet{fukugita96a} to obtain 
\begin{equation}
M_{u^\ast{\odot}}=6.39 \quad;\quad
M_{g^\ast{\odot}}=5.07 \quad;\quad
M_{r^\ast{\odot}}=4.62 \quad;\quad
M_{i^\ast{\odot}}=4.52 \quad;\quad
M_{z^\ast{\odot}}=4.48.
\end{equation}
These numbers agree well with those predicted from theoretical models
of G2V stars (\citealt{lenz98a}; \citealt{fan99a}). There still may be
errors at the level of 5\% in the photometric calibration for the data
analyzed here; furthermore, the color transformations are somewhat
uncertain, perhaps also by 5\%. For this reason we add in quadrature a
7\% systematic uncertainty to the errors listed for the luminosity
density in solar units. The luminosity densities determined with this
analysis are consistent with those determined by Yasuda 
{\it et al.}\ (in preparation)
based on the bright number counts in a similar SDSS data set along
with the parameters $M_\ast$ and $\alpha$ of the Schechter function
measured here, providing a check on our method of determining the
normalization.

\subsection{Redshift Distribution of SDSS Galaxies}

\begin{figure}
\figurenum{\fignum}
\plotone{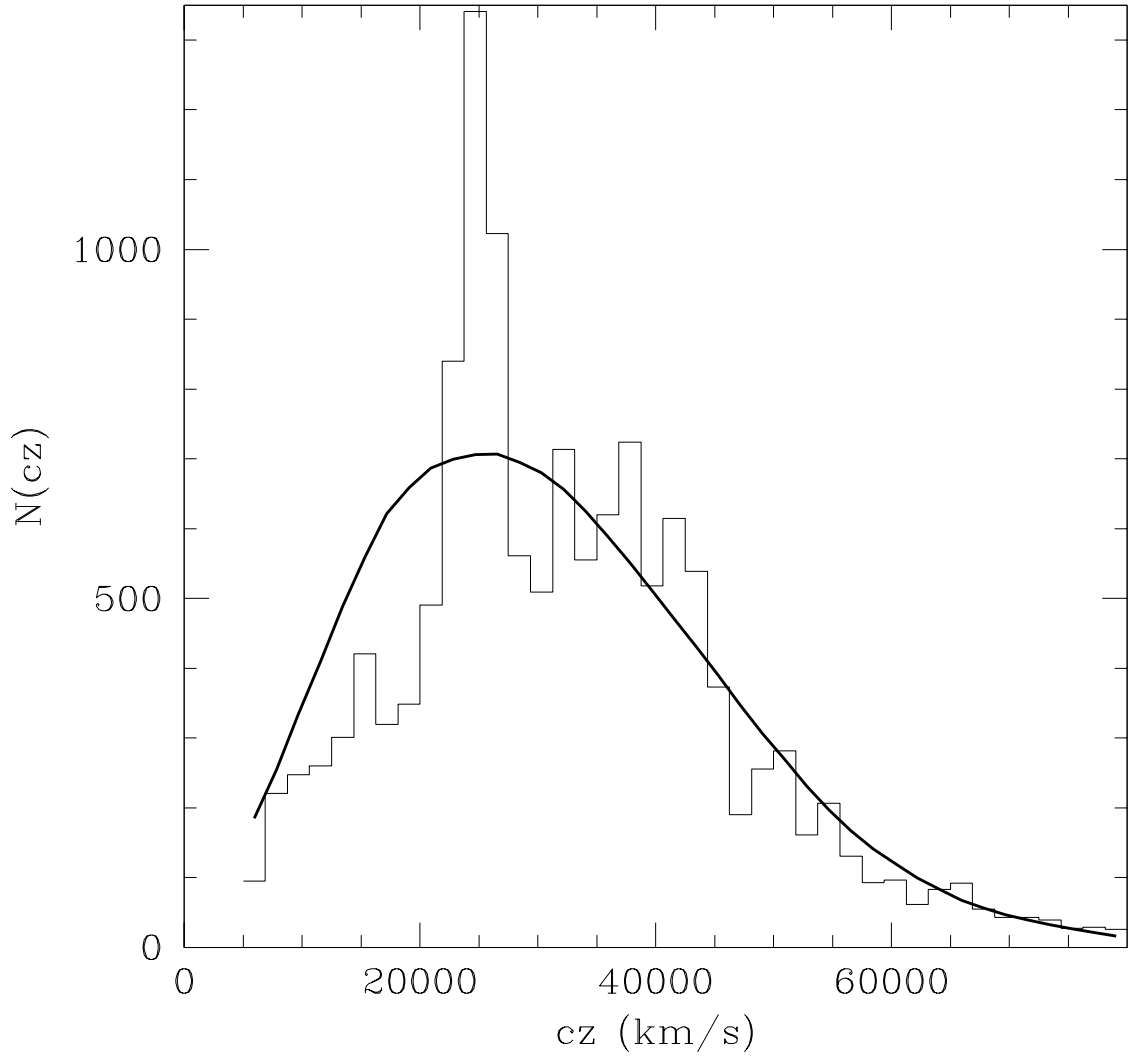}
\caption{\label{rh} Redshift distribution of SDSS galaxies (histogram)
compared to the average distribution expected given the $r^\ast$
luminosity function and the flux limits (using $K$-corrections for a
``typical'' galaxy color of $g^\ast-r^\ast=0.65$). The redshift distribution
is reasonable, although large-scale structure is obvious.}
\end{figure}

Here we compare the observed redshift distribution to the expected
redshift distribution of a homogeneous galaxy sample based on our
luminosity function and the flux limits. We calculate the expected
number of galaxies using a ``typical'' galaxy color of
$g^\ast-r^\ast=0.65$ in order to calculate the $K$-correction. One can
refine this analysis using the joint color-luminosity relation we
calculate below in Section \ref{color}, and it is important to do so
when calculating large-scale structure statistics.  However, for
current purposes picking a typical color will suffice. To calculate
the observed redshift distribution we weight each galaxy according to
its ``collision weight,'' $w_c$, described in Section \ref{sampling}.
Figure \ref{rh} compares the expected counts to the observed
counts. Clearly large-scale structure is noticeable in an area of sky
this small, but overall the redshift distribution appears consistent
with that predicted by the luminosity function. Zehavi {\it et al.}
(in preparation) are measuring the large-scale structure in this
sample and find that it is generally consistent with what has been
found in previous surveys. We should emphasize again that the
likelihood method used here to calculate the shape of the luminosity
function is insensitive to the effects of large-scale structure, to
the extent that the luminosity function does not depend on
environment.

\begin{figure}
\figurenum{\fignum}
\plotone{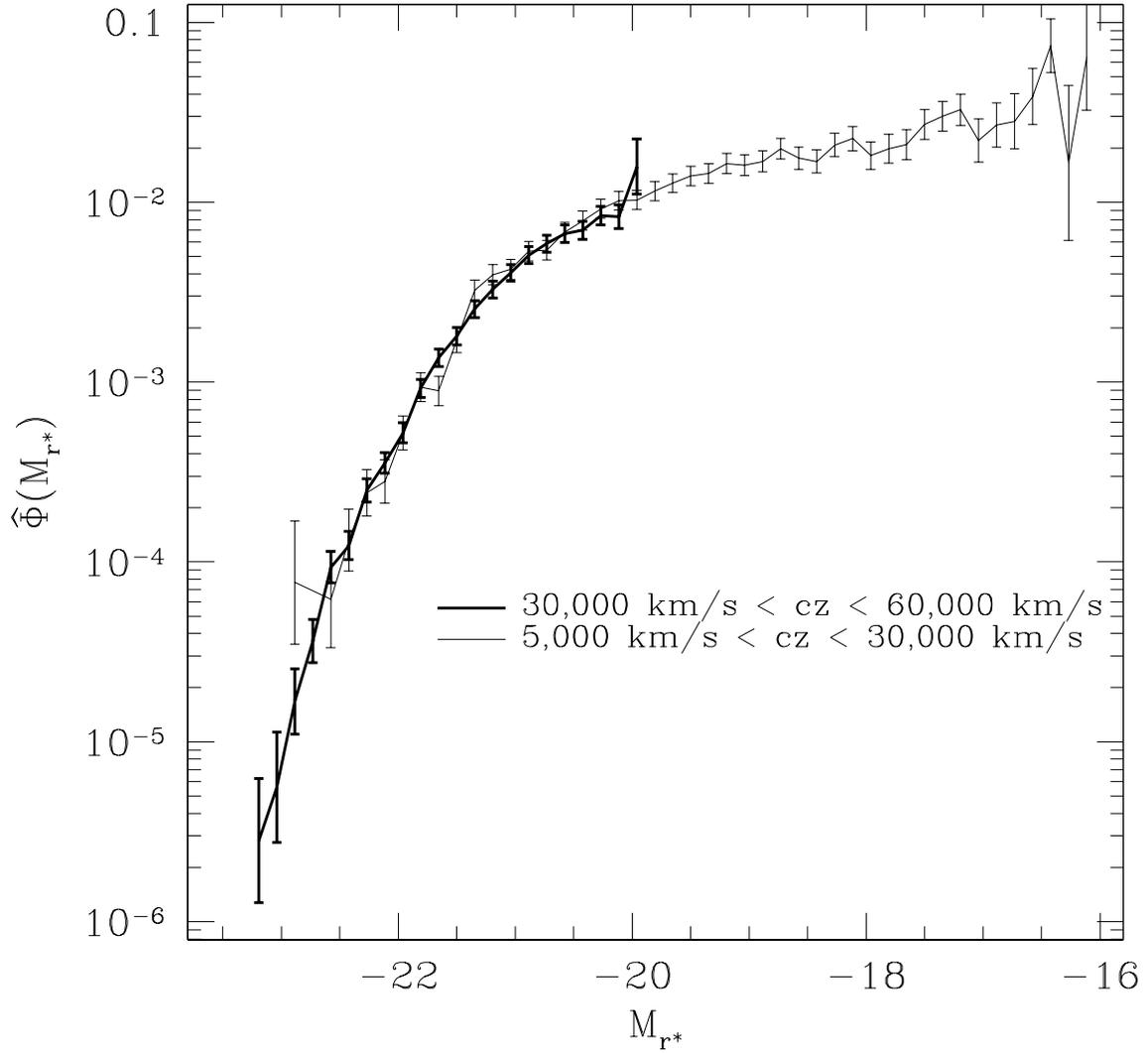}
\caption{\label{hilo} Comparison of the luminosity function in $r^\ast$
determined using samples above and below $cz=30,000$ km s$^{-1}$. In the
region in which the two luminosity functions overlap they are in good
agreement, indicating that the redshift dependence of the selection
function is purely due to the flux limits of the survey. More precise
tests of this nature can be made when larger areas of the sky (which
are less affected by large-scale structure) become available.}
\end{figure}

To reassure ourselves that the method of targeting galaxies does not
introduce redshift-dependent selection, apart from the effect of the
flux limits, we divide the set of galaxies into two subsamples above
and below $cz = 30,000$ km s$^{-1}$, and calculate the luminosity
function separately for these two samples. Figure
\ref{hilo} compares the results. The two luminosity
functions agree in the absolute magnitude range in which they
overlap. More precise comparisons of the normalization of luminosity
functions at high and low redshift will be possible once more area of
the sky is available for analysis.

\subsection{Surface Brightness Selection Effects and the Correlation
of Luminosity with Surface Brightness}
\label{sbdep}

Our assertion that we are measuring the galaxy luminosity function
implicitly assumes that our sample is truly flux-limited, and that
surface brightness selection effects do not influence the luminosity
function at a significant level. As described in Section \ref{target},
the commissioning observations incorporate an explicit surface
brightness cut at $\mu_{1/2,r^\ast}=23.5$. In the apparent magnitude
range $14.5<r^\ast<17.6$ used here, we expect the SDSS imaging to
detect essentially all galaxies above this surface brightness
threshold, so there should be no {\it additional} selection effects
associated with failure to detect low surface brightness galaxies in
the first place. The half-light radius of a galaxy with apparent
magnitude $m$ and half-light surface brightness $\mu_{1/2}$ is
$r_{1/2} = 6.0'' \times 10^{0.2(\mu_{1/2}-23.5)+0.2(17.6 -m)}$. Near
the magnitude limit $r^\ast=17.6$ and the surface brightness limit
$\mu_{1/2}=23.5$, this size is well below the $\sim 50''$ scale used
to define the background for sky subtraction. At $r^\ast = 14.5$, the
half-light radius is $\sim 25''$, so sky subtraction and deblending
failures could have some impact on detection, but only a tiny fraction
of the galaxy targets would be influenced, many of which would be too
close in redshift to be included in the sample used here for the
luminosity function (Yasuda {\it et al.}, in preparation, discuss this
issue in more detail). At the depth of the SDSS imaging a galaxy with
$r^\ast=17.6$ and $\mu_{1/2,r^\ast}=23.5$ has mean $S/N \approx 50$
within $r_{1/2}$, so detections have high statistical significance. In
principle, one can also miss high surface brightness galaxies by
mis-classifying them as stars, but Figure \ref{psfmodel} shows that
the distribution of the quantity $m_{\mathrm{PSF}}-m_{\mathrm{model}}$
used for star-galaxy separation has fallen off well before reaching
the classification threshold. The number of high surface brightness
galaxies missed is therefore a tiny fraction of the full sample; if we
choose a higher cut-off for $m_{\mathrm{PSF}}-m_{\mathrm{model}}$
which excludes 1\% of the canonical galaxy sample, the effect on the
luminosity and surface-brightness distributions is negligible.

The $\mu_{1/2,r^\ast}=23.5$ surface brightness limit itself only
excludes about 0.5\% of galaxies within our flux limits.  However, the
well-known relationship between surface brightness and luminosity
(\citealt{lilly98a}; \citealt{driver99a}; \citealt{dejong00a}),
confirmed below, suggests that the excluded galaxies will
preferentially be of low luminosity. Since these underluminous
galaxies are observable only within a rather small, nearby volume,
they are given disproportionate statistical weight when the luminosity
function is calculated, and thus even a small number of missing
objects may significantly bias our estimate of the luminosity function
at low luminosities. This effect is one reason why the full SDSS
sample will have a fainter surface brightness limit than 23.5, as
mentioned in Section \ref{target} above.

\begin{figure}
\figurenum{\fignum}
\plotone{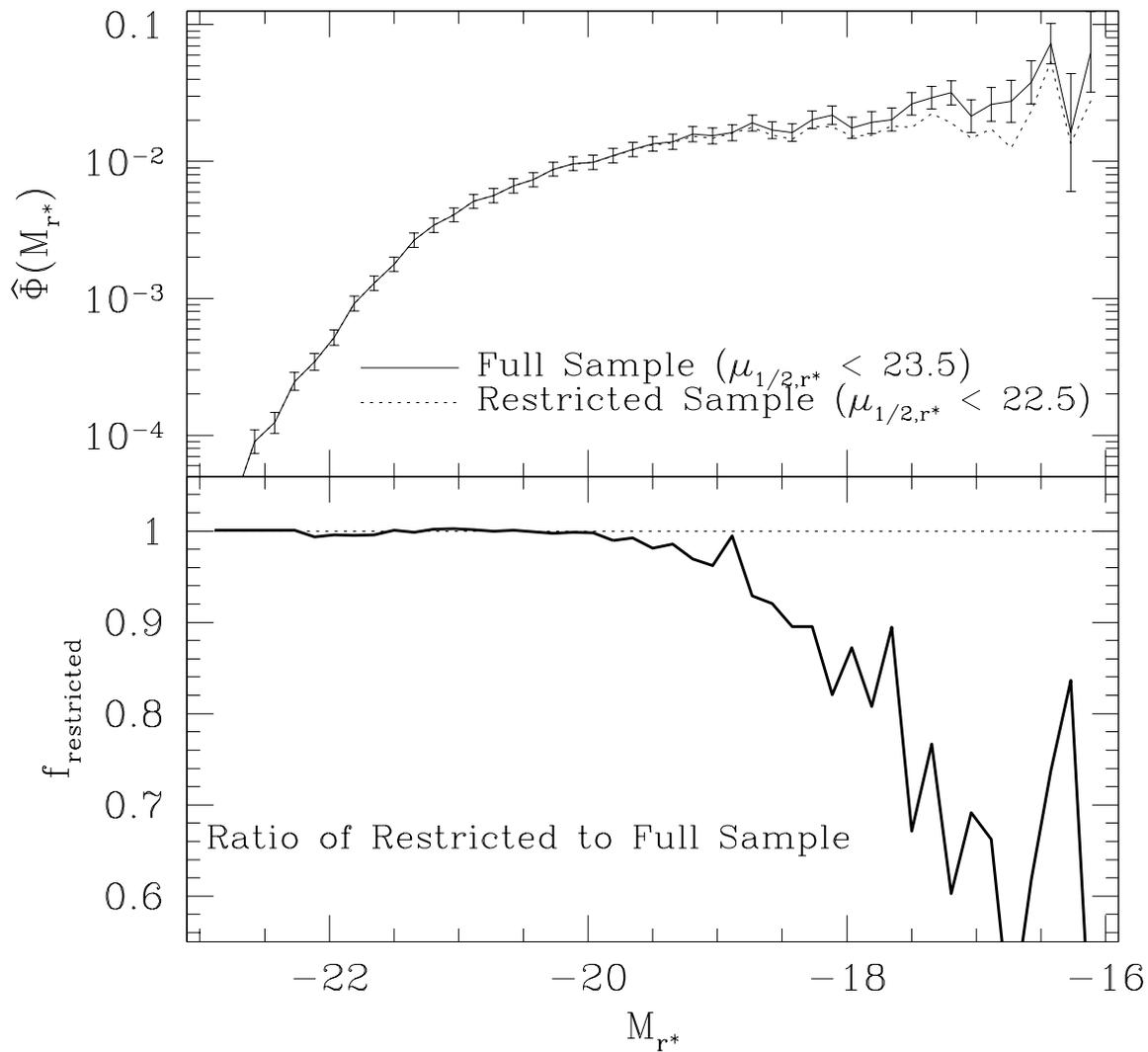}
\caption{\label{surfbright} Luminosity
function of SDSS galaxies in the $r^\ast$ band for the full sample,
limited at a surface-brightness of $\mu_{1/2,r^\ast} < 23.5$, and for
a restricted sample with $\mu_{1/2,r^\ast} < 22.5$. The higher surface
brightness threshold excludes about 1.5\% of the galaxies in the full
sample. The top panel shows both luminosity functions, showing that
differences between them appear at $M_{r^\ast}>-19$.  The bottom panel
shows the ratio of the two luminosity functions.}
\end{figure}

A simple way to investigate this effect is to raise the surface
brightness threshold to $\mu_{1/2,r^\ast} = 22.5$.  This restriction
excludes an additional 1.5\% of the galaxies within our flux limits.
Figure \ref{surfbright} shows the luminosity function of the full and
restricted samples in the top panel, and the ratio of the luminosity
functions in the bottom panel.
Significant numbers of galaxies are missed at $M_{r^\ast}>-19$ in the
restricted sample. In a Schechter function fit to the samples, the
slope $\alpha$ is shallower by about 0.08 in the restricted sample,
and the influence on $\alpha$ is approximately the same in the other
bands.  For galaxies in the absolute magnitude range
$-23.5<M_{r^\ast}<-16$, the restricted sample has about 5\% lower
luminosity density than does the full sample, as measured from the
non-parametric fits. If one extrapolates the fitted Schechter
functions to all luminosities, the difference is still only
5\%. However, if the low-luminosity slope $\alpha$ steepens
considerably below $M>-16$, then the contribution of low surface
brightness galaxies to the luminosity could be larger.  Because the
objects with $\mu_{1/2,r^\ast}>23.5$ have a surface number density
that is about 0.5\% of that of galaxies in our sample, these effects
represent a firm upper limit on the influence of our surface
brightness threshold on the luminosity function in the range of
absolute magnitudes considered here.

\begin{figure}
\figurenum{\fignum}
\plotone{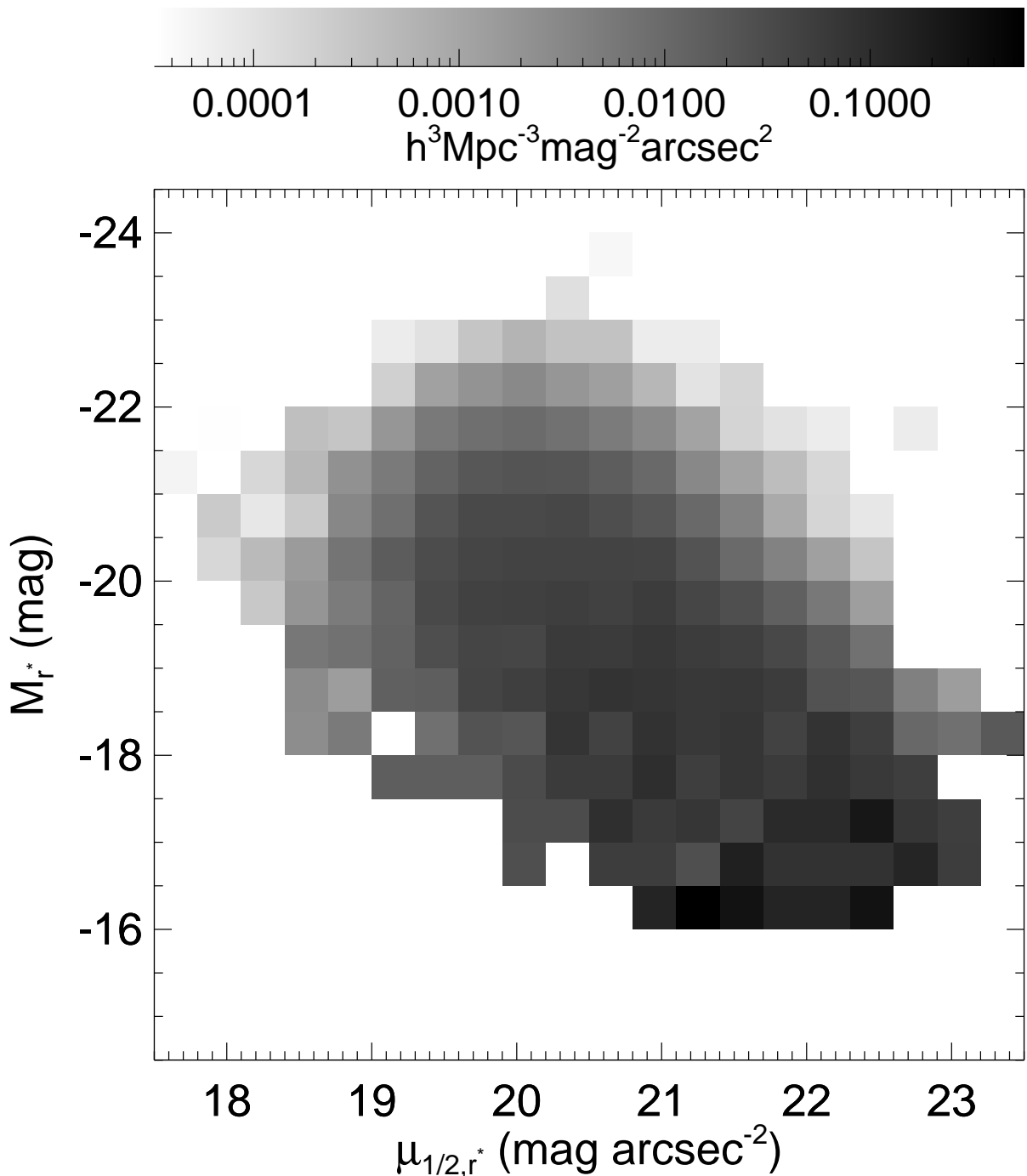}
\caption{\label{lsbf} The logarithmic greyscale represents
$\hat\Phi(M_{r^\ast},\mu_{1/2,r^\ast})$ in $r^\ast$ for SDSS galaxies,
in units of galaxies per $h^{-3}$ Mpc$^{3}$ per unit magnitude per
unit surface brightness, calculated using the method of
\citet{efstathiou88a}. The translation of the greyscale to these units
is given in the bar at the top of the figure. The data sample has an
explicit half-light surface brightness cut at
$\mu_{1/2,r^\ast}=23.5$. The strong correlation between surface
brightness and luminosity is apparent. It is clear why many of the
objects with $\mu_{1/2,r^\ast}>22.5$ are low luminosity, as shown in
Figure \ref{surfbright}. }
\end{figure}

\begin{figure}
\figurenum{\fignum}
\plotone{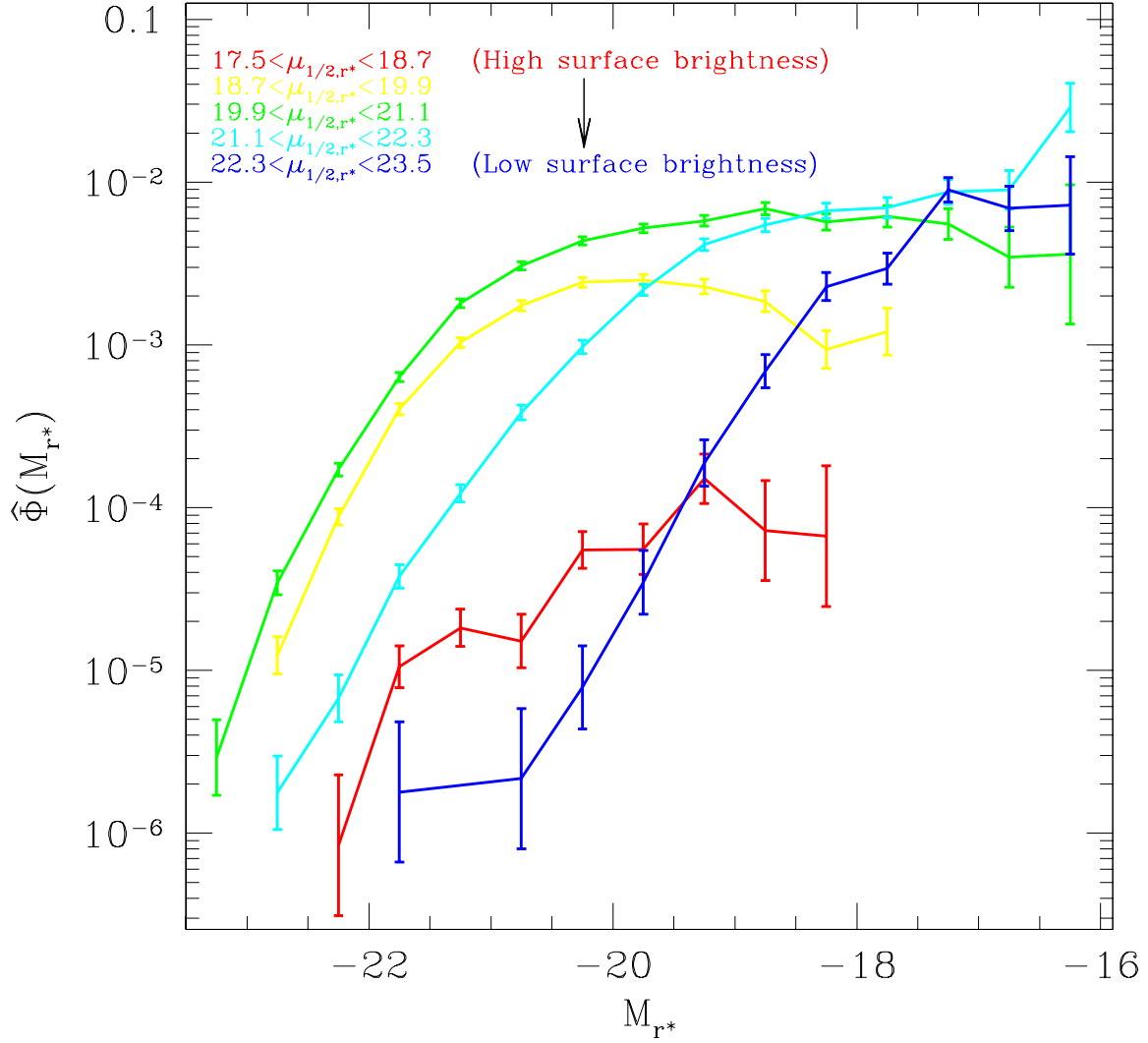}
\caption{\label{lsbf_M} Luminosity function of SDSS galaxies in the
$r^\ast$ band, in bins of half-light surface brightness, in units of
galaxies per $h^{-3}$ Mpc$^{3}$ per unit magnitude. Each curve shows
the luminosity function for a range of surface-brightnesses.
As labeled, the curves are colored blue, cyan, green, yellow, and red
from high to low surface brightness.  
The steep faint-end slope for
low surface-brightness galaxies is apparent, as well as the fact that
the high surface-brightness galaxies tend to be luminous.}
\end{figure}

\begin{figure}
\figurenum{\fignum}
\plotone{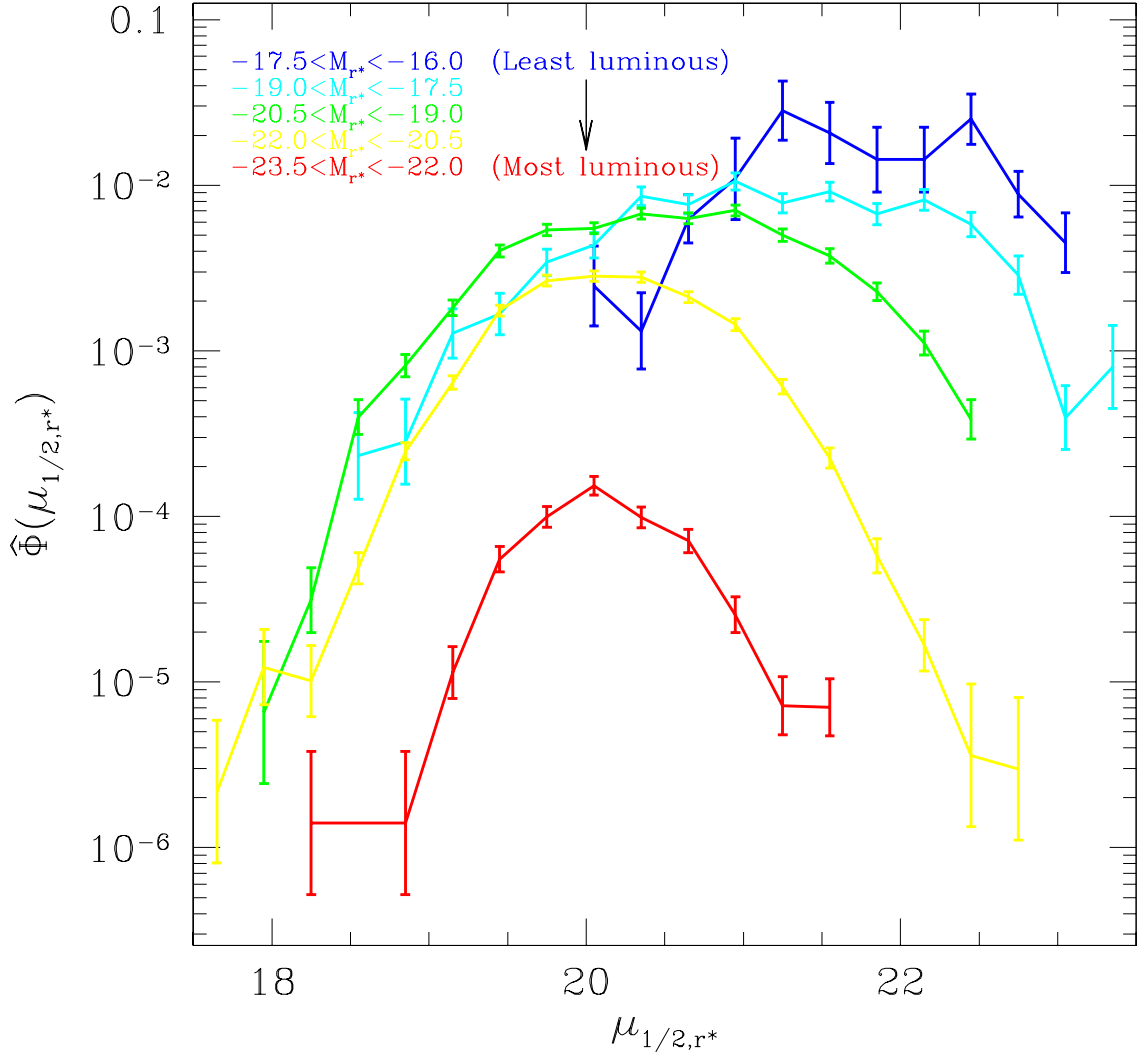}
\caption{\label{lsbf_sb} Half-light surface-brightness function of
SDSS galaxies in the $r^\ast$ band, in bins of galaxy luminosity, in
units of galaxies per $h^{-3}$ Mpc$^{3}$ per unit surface
brightness. Each curve shows the surface brightness function for a
range of $M_{r^\ast}$.
As labeled, the curves are colored blue, cyan, green, yellow, and red
from low to high luminosity.  The luminous galaxies tend to have a
characteristic half-light surface brightness, while low luminosity
galaxies have a broad distribution that extends to low
surface-brightness.}
\end{figure}

The joint distribution of surface brightness and luminosity is a
valuable diagnostic for understanding surface brightness selection
effects and an important quantitative characterization of the galaxy
population in its own right (see, {\it e.g.} \citealt{driver00a} and
references therein). To define this distribution, we take
$\mu_{1/2,r^\ast}$ (corrected for cosmological surface brightness
dimming) as the measure of surface brightness, and use the
two-dimensional version of the non-parametric method of
\citet{efstathiou88a} to calculate the joint distribution of
$\mu_{1/2,r^\ast}$ and $M_{r^\ast}$, taking into account the apparent
magnitude and surface brightness limits of the sample. The results of
this fit are shown in Figure \ref{lsbf}. Figures \ref{lsbf_M} and
\ref{lsbf_sb} show slices parallel to the axes of this two-dimensional
plane: the luminosity function for several ranges of intrinsic surface
brightness (Figure \ref{lsbf_M}) and the surface brightness
distribution for several ranges of absolute magnitude (Figure
\ref{lsbf_sb}). Note that we have not corrected the values of
$\mu_{1/2,r^\ast}$ for the effects of seeing, so the surface
brightness of smaller galaxies is systematically underestimated.

Figures \ref{lsbf}---\ref{lsbf_sb} demonstrate a number of interesting
points.  Galaxies with $M_{r^\ast}< -19$ show a clear peak in their
surface brightness distribution, with typical $\mu_{1/2,r^\ast}
\approx 20$.  This result is in rough agreement with the
previous results of \citet{freeman70a} and \citet{courteau96a}, although
those authors measured the central, rather than the half-light,
surface brightness, and concerned themselves only with disk galaxies.  In
addition, galaxy surface brightness is clearly correlated with
luminosity, in agreement with the results of, {\it e.g.},
\citet{dejong00a}. For lower luminosities, the typical surface
brightness is lower, and the distribution of surface brightness is
broader (\citealt{phillipps86a}). For lower surface brightness, the
galaxy luminosity function becomes steeper. The intrinsic surface
brightness distribution of galaxies with $M_{r^\ast}<-19$ drops off
well before our threshold of $\mu_{1/2,r^\ast} = 23.5$, providing
further evidence that our luminosity function is unaffected by surface
brightness selection in this magnitude range. For galaxies less
luminous than $M_{r^\ast} \approx -17.5$, the surface brightness
distribution does decline before hitting our threshold, but not in as
definitive a manner as it does for more luminous galaxies; thus, our
galaxy density in this regime might be underestimated, and we cannot
rule out an upturn in the luminosity function at these
magnitudes. However, the luminosity density for a Schechter luminosity
function with $\alpha>-2$ is dominated by galaxies with $M\approx
M_\ast$. Unless there is a radical change in the shape of the
luminosity function at $M_{r^\ast}>-17.5$, we can safely say that low
surface brightness galaxies make a relatively small contribution to
the luminosity density of the universe. All of these conclusions are
in broad agreement with most recent studies (\citealt{impey97a}).

There is great potential for further studies of galaxy surface
brightness and its correlation with other properties using the
SDSS. As a result of improvements in the photometric data reduction
software, the main survey (now underway) will adopt a fainter surface
brightness threshold, $\mu_{1/2,r^\ast} = 24.5$. This change will
improve the measurement of the luminosity function and the bivariate
luminosity-surface brightness distribution at lower luminosities. With
the aid of simulated data, it should be possible to correct estimated
$\mu_{1/2}$ values for seeing effects. With a larger data sample, it
will be possible to analyze joint distributions of surface brightness
with a larger set of other galaxy parameters, such as color and
morphology, using all five bandpasses. Finally, the SDSS imaging will
provide an excellent data set for finding candidate galaxies far below
the $\mu_{1/2,r^\ast} = 24.5$ threshold, even if their redshifts are
not obtained by the SDSS spectroscopic survey. Follow-up efforts in
optical and 21 cm and correlation with HI radio surveys such as HIPASS
({\it e.g.}, \citealt{banks99a}) will greatly improve the
characterization of the low surface brightness galaxy population.

\subsection{Dependence of Luminosity on Color}
\label{color}

It is well-known that galaxies of different intrinsic colors have
quite different luminosity functions. We explore this issue here in a
simple way, by calculating the joint distribution of luminosity and
the intrinsic $g^\ast-r^\ast$ color. We determine the intrinsic
$g^\ast-r^\ast$ color for a galaxy at a particular redshift by using
the observed $g^\ast-r^\ast$ color to interpolate between the
predicted colors at that redshift for various galaxy types in the
results of \citet{fukugita95a}. There are only five templates between
which we are interpolating, which may affect the accuracy of our color
estimates (typical color corrections to rest frame are about 0.25
magnitudes, so our accuracy is considerably better than that).

\begin{figure}
\figurenum{\fignum}
\plotone{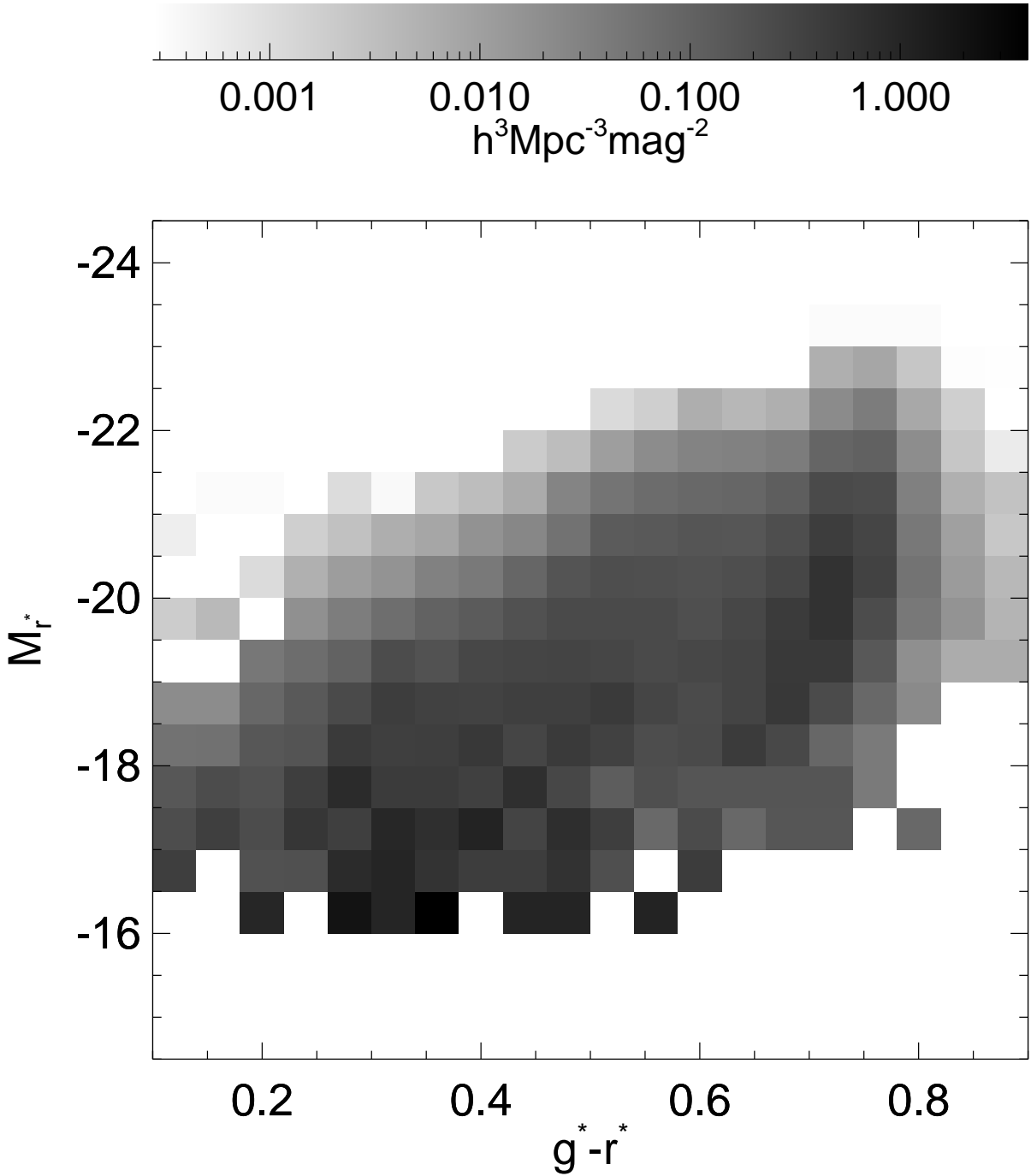}
\caption{\label{lgmrf} Same as Figure \ref{lsbf}, for the joint
luminosity-color distribution $\hat\Phi(M_{r^\ast},g^\ast-r^\ast)$ in
$r^\ast$ for SDSS galaxies, now in units of galaxies per $h^{-3}$
Mpc$^{3}$ per unit magnitude per unit color.  
The intrinsic $g^\ast-r^\ast$ colors are inferred using the observed
$g^\ast-r^\ast$ color, the measured redshift, and the results of
\citet{fukugita95a}. The ``E/S0'' ridge at $g^\ast-r^\ast \approx
0.75$ is apparent, as is the strong correlation between
luminosity and color.}
\end{figure}

\begin{figure}
\figurenum{\fignum}
\plotone{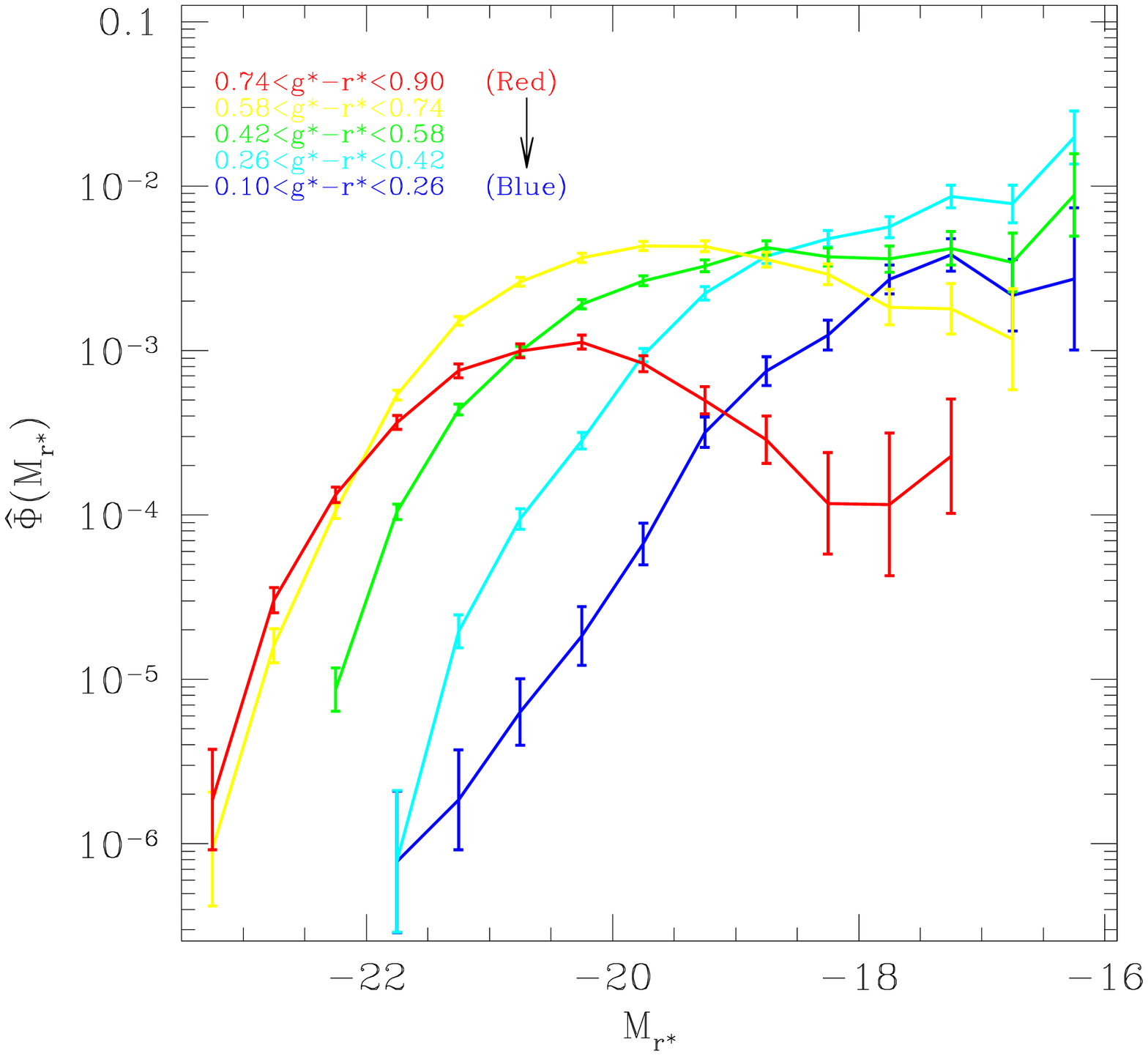}
\caption{\label{lgmrf_M} Same as Figure \ref{lsbf_M}, only now showing
the luminosity function in bins of intrinsic $g^\ast-r^\ast$
color, in units of galaxies per $h^{-3}$ Mpc$^{3}$ per unit magnitude.
As labeled, the curves are colored blue, cyan, green, yellow, and red
from bluest to redder.  The intrinsic color is determined using the
observed color, the measured redshift, and the results of
\citet{fukugita95a}. Errors in the calibration used for this data
probably broaden this distribution slightly, by about 0.05
magnitudes.}
\end{figure}

\begin{figure}
\figurenum{\fignum}
\plotone{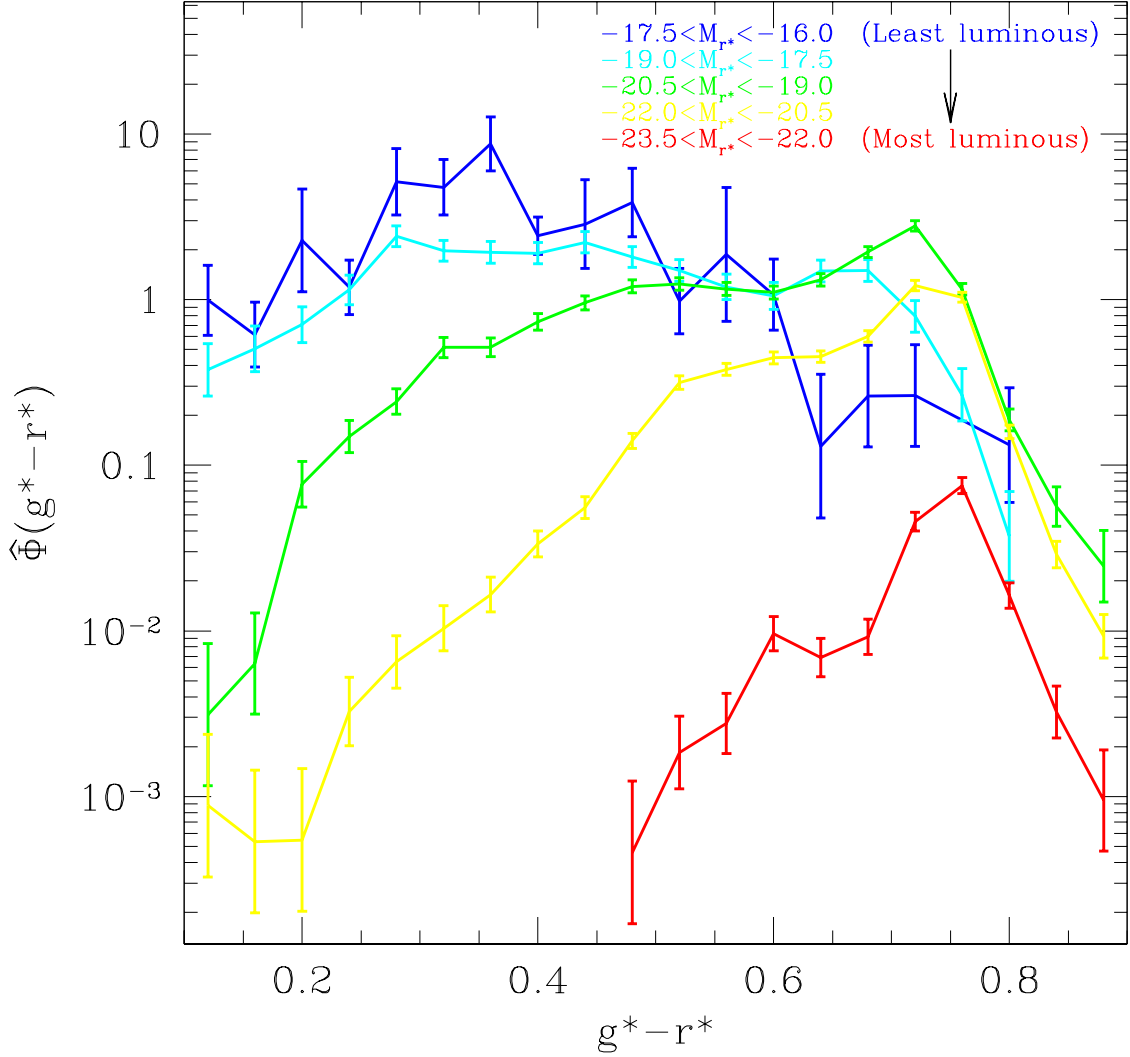}
\caption{\label{lgmrf_gmr} Same as Figure \ref{lsbf_sb}, only now
showing the distribution of intrinsic $g^\ast-r^\ast$ color for
several ranges of $r^\ast$ luminosity, in units of galaxies per
$h^{-3}$ Mpc$^{3}$ per unit color. 
As labeled, the curves are colored blue, cyan, green, yellow, and red
from low to high luminosity.  As in Figure \ref{lgmrf}, the ``E/S0''
ridge at $g^\ast-r^\ast \approx 0.75$ is apparent for high
luminosities, while for low luminosities, the $g^\ast-r^\ast$
distribution is nearly flat. }
\end{figure}

The resulting distribution, $\hat\Phi(M_{r^\ast}, (g^\ast-r^\ast)_0)$
is shown in Figure \ref{lgmrf}.  This figure shows the ``E/S0'' ridge
at $(g^\ast-r^\ast)_0 \sim 0.75$, which is well-known in the context
of galaxy clusters, as well as the dependence of color on
luminosity. As in the case of surface brightness, we show the
luminosity function for several ranges of color in Figure
\ref{lgmrf_M}, and the color distribution as a function of luminosity
in Figure
\ref{lgmrf_gmr}. The ridge at $(g^\ast-r^\ast)_0 \sim 0.75$ is
apparent at high luminosities in Figure \ref{lgmrf_gmr}, as is the
relative flatness of the color distribution at low luminosity. 




A better examination of the dependence of galaxy color on luminosity
will be possible once the photometric calibration of the data is
complete (although the calibrations used here are thought to be
accurate to 0.05 magnitudes). Once the final calibration is available,
another improvement will be to use the broad-band imaging (and
possibly the spectra) to determine the $K$-corrections from the data
itself. Methods developed for photometric redshifts, which can use the
broad-band colors to reconstruct the typical spectral energy
distributions of galaxies ({\it e.g.},
\citealt{csabai00a}), can be brought to bear on this problem and
will resolve the difficulties mentioned above associated with the
small number of templates used for interpolation.

\subsection{Dependence of Luminosity on Morphology}
\label{morphology}

Finally, let us consider the dependence of luminosity on the
morphology of galaxies. Although there are numerous ways to quantify
galaxy morphology, we choose here for simplicity to use the
``concentration index,'' defined to be $c=r_{90}/r_{50}$, the ratio of
the radii containing 90\% and 50\% of the Petrosian flux. We will
actually study the inverse concentration index $1/c$ here, because it
has the property of being between zero and one. The concentration
index is high ($c \approx 3.3$, or $1/c\approx 0.3$) for pure de
Vaucouleurs profile galaxies and low ($c \approx 2.3$, or $1/c
\approx 0.43$) for pure exponential profiles.\footnote{As mentioned 
in Section \ref{petro}, $r_{50}$ and $r_{90}$ are
defined with respect to the {\it Petrosian} flux; the inverse concentration
parameter for de Vaucouleurs profiles when $r_{50}$ and
$r_{90}$ are measured with respect to the {\it total} flux is about
$1/c \approx 0.18$.} 
It has been shown to correlate well with visual morphological
classification for bright galaxies (Shimasaku {\it et al.}, in
preparation; Strateva {\it et al.}, in preparation). 

\begin{figure}
\figurenum{\fignum}
\epsscale{0.9}
\plotone{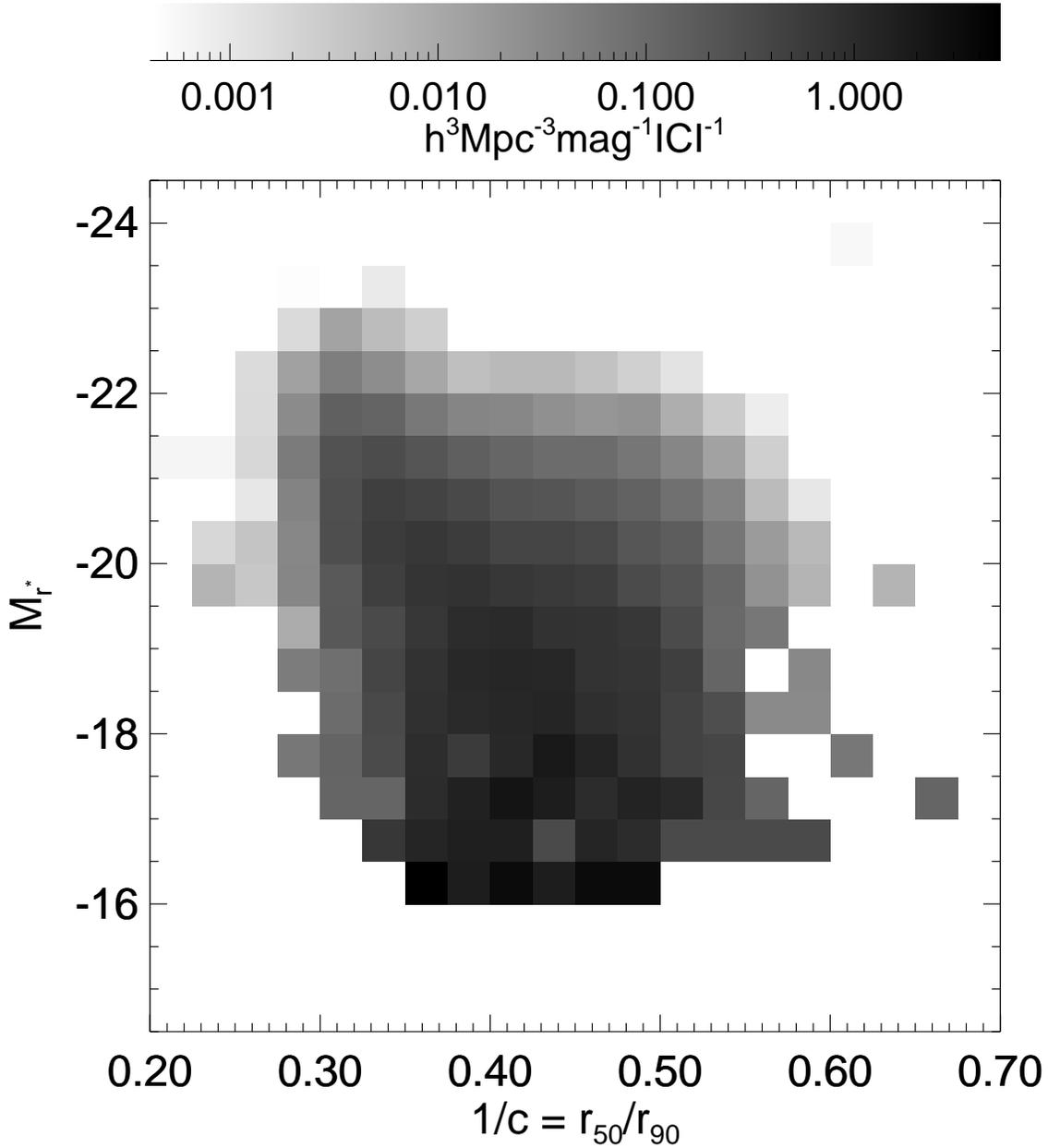}
\caption{\label{lcrf} Same as Figure \ref{lsbf}, for the joint
distribution of luminosity and inverse concentration index
$\hat\Phi(M_{r^\ast},1/c)$ in $r^\ast$ for SDSS galaxies, in units of
galaxies per $h^{-3}$ Mpc$^{3}$ per unit magnitude per unit inverse
concentration index (ICI).  The concentration index is defined to be
$c=r_{90}/r_{50}$. Again, the galaxies with a low inverse
concentration index ({\it i.e.}, those galaxies with profiles closer
to de Vaucouleurs) tend to be brighter. Note, however, the abundance
of high inverse concentration index galaxies ($1/c >0.45$). Since
exponential disks should have inverse concentration indices around
0.43, these galaxies either have much broader profiles than
exponential, or are being affected by seeing and noise in the
measurement. Nevertheless, the trend of luminosity with morphology is
clear. }
\end{figure}

\begin{figure}
\figurenum{\fignum}
\epsscale{1.0}
\plotone{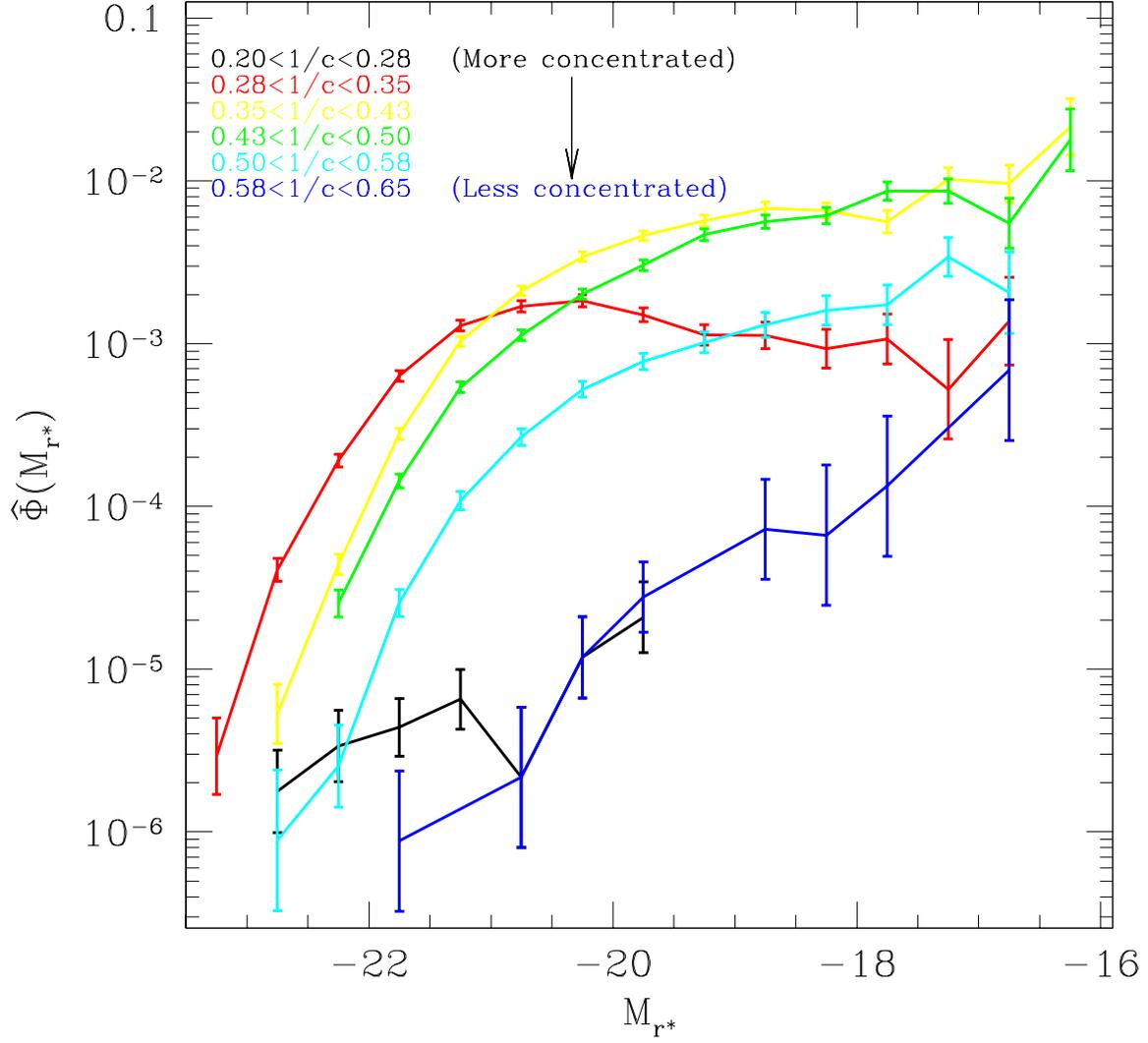}
\caption{\label{lcrf_M} Same as Figure \ref{lsbf_M}, only now showing
the luminosity function in bins of inverse concentration index
$1/c=r_{50}/r_{90}$, in units of galaxies per $h^{-3}$ Mpc$^{3}$ per
unit magnitude.  As labeled, the curves are colored blue, cyan, green,
yellow, red, and black from least to most concentrated.}
\end{figure}

\begin{figure}
\figurenum{\fignum}
\plotone{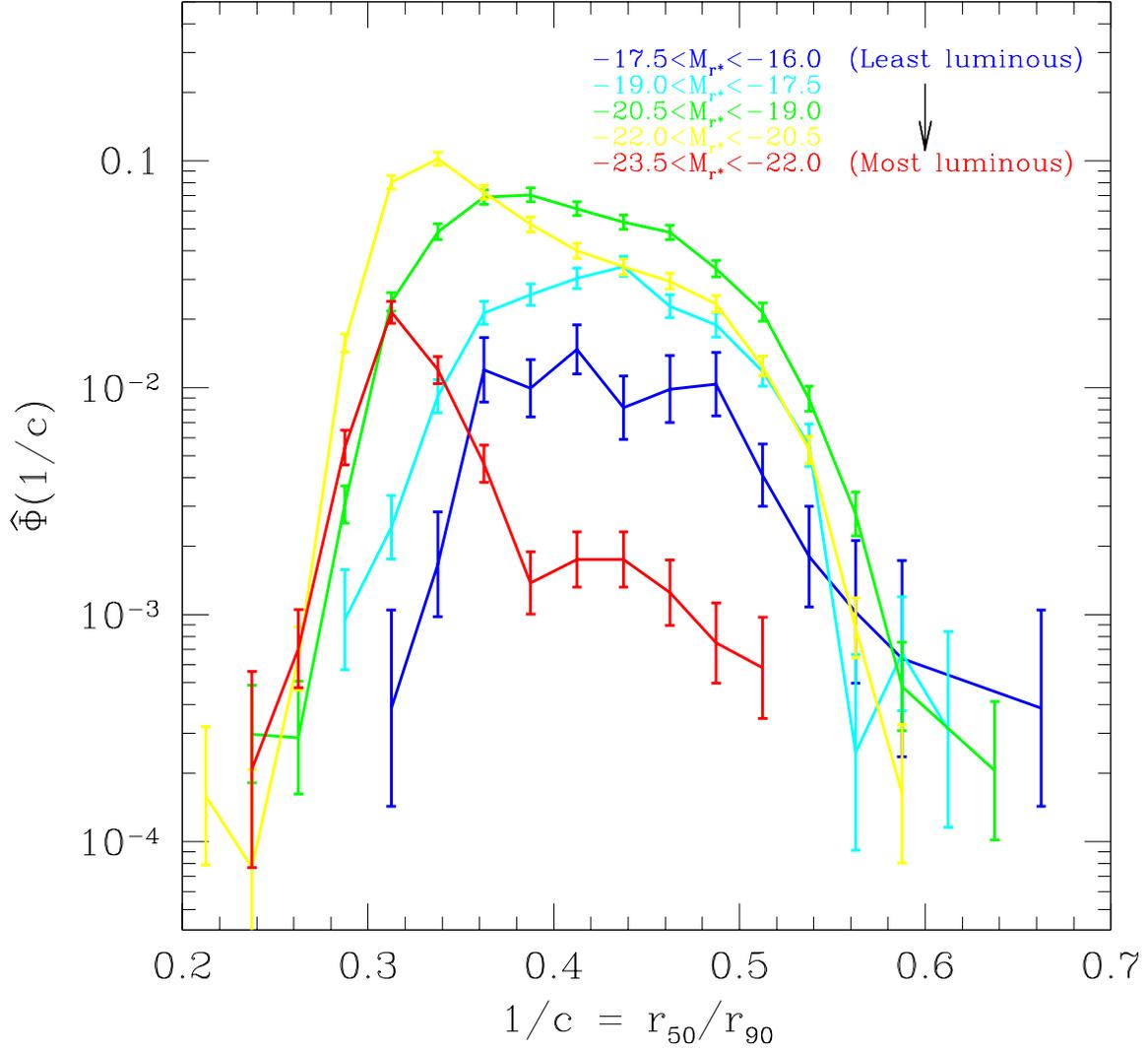}
\caption{\label{lcrf_cr} Same as Figure \ref{lsbf_sb}, only now
showing the distribution of inverse concentration index
$1/c=r_{50}/r_{90}$ for several ranges of $r^\ast$ luminosity, in
units of galaxies per $h^{-3}$ Mpc$^{3}$ per unit inverse
concentration index.
As labeled, the curves are colored blue, cyan, green, yellow, and red
from low to high luminosity. }
\end{figure}

Figure \ref{lcrf} shows the joint distribution of luminosity and
inverse concentration index. As one might expect, the high luminosity
galaxies tend to have high concentration index, and thus to have
profiles which are closer to de Vaucouleurs profiles.  As before, we
slice this two-dimensional plane to get the luminosity function for
several ranges of the inverse concentration index in Figure
\ref{lcrf_M}, and the distribution of the inverse concentration index
for several ranges of luminosity in Figure \ref{lcrf_cr}.

Bear in mind that the concentration index is not the only, and perhaps
not the best, measure of morphology. For example, some of the inverse
concentration indices measured may be artificially affected by the
effects of seeing. The inverse concentration index for a realistic PSF
measured from the survey is around $1/c \sim 0.5$, so exponential
disks and de Vaucouleurs profiles that are marginally resolved will
tend to have their inverse concentration indices increased towards
this value.
Furthermore, this measure of morphology is based only on the radial
profile in $r^\ast$, and thus does not use the information on color
gradients, shape, and galaxy substructure that is available in the
imaging. Developing measures of morphology that use the full
information from the galaxy images and that are independent of the
effects of seeing is one of the ongoing projects within the SDSS
collaboration.

\section{Comparison of Petrosian Magnitudes to Isophotal Magnitudes}
\label{isopetro}

As discussed in Section \ref{target}, we have selected spectroscopic
targets and calculated galaxy luminosities using Petrosian magnitudes,
so that the fraction of a galaxy's light that is measured is
independent of the amplitude of its surface brightness profile and
independent of cosmological redshift dimming or Galactic extinction.
In order to compare our luminosity function results to those of
previous surveys, most of which selected galaxies according to some
variant of an isophotal magnitude, we must understand how the
difference between Petrosian and isophotal magnitudes affects the
determination of the luminosity function.

We can calculate an isophotal magnitude for each galaxy in our sample,
by integrating the azimuthally-averaged radial profile of each galaxy,
which is available as a standard output of {\tt photo} (Lupton {\it et
al.}, in preparation), out to a chosen isophote.  These magnitudes do
not correspond exactly to the common definition of isophotal
magnitudes, which typically are not azimuthally-averaged. However, as
we show in the next section by comparing the SDSS results to those of
other surveys, the two definitions are probably nearly equivalent, at
least for luminosity statistics.

\begin{figure}
\figurenum{\fignum}
\plotone{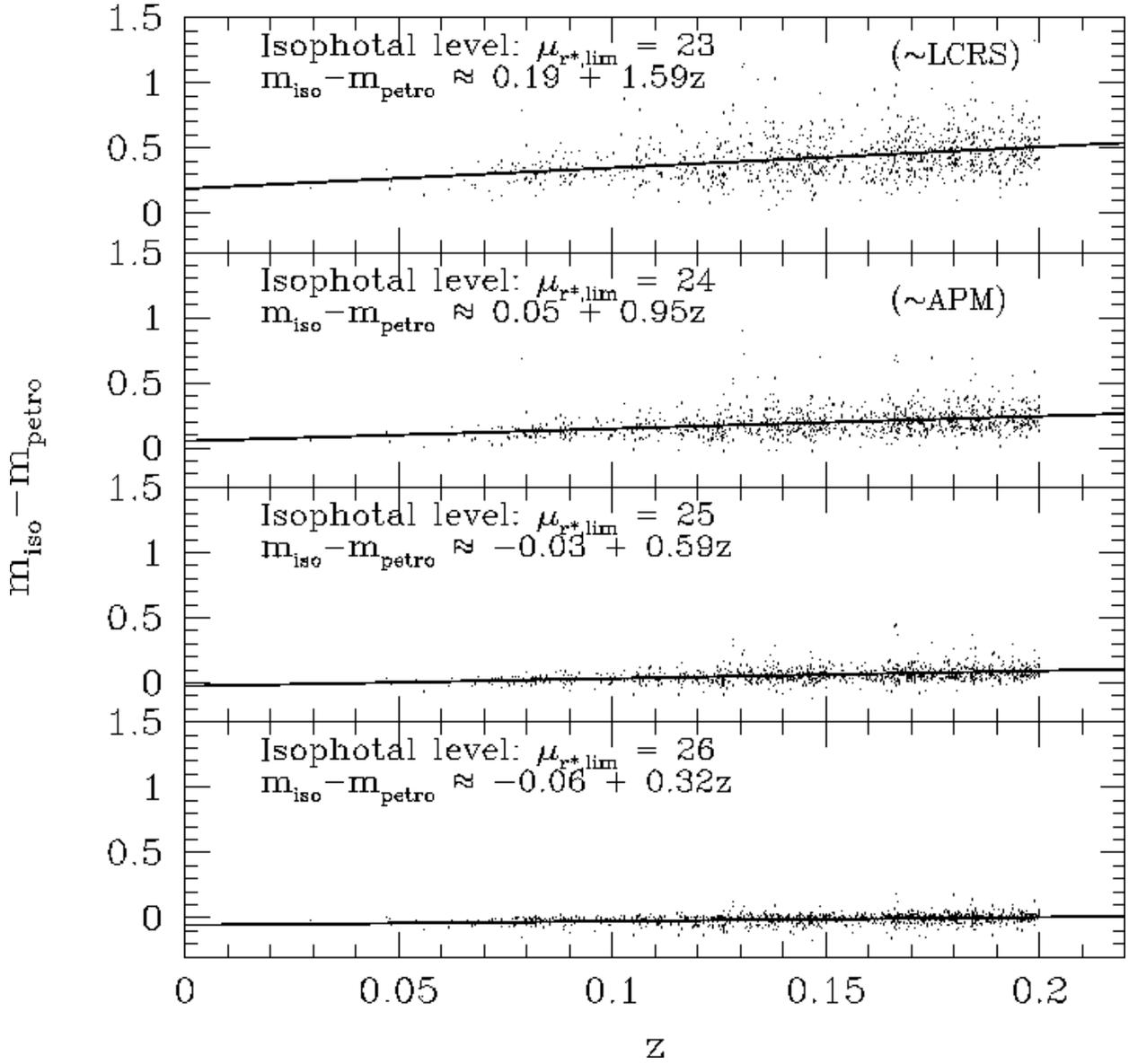}
\caption{\label{isomag_z_vol} 
Comparison of isophotal and Petrosian magnitudes as a function of
redshift in the $r^\ast$ band for several choices of the isophotal
limit.  We only show galaxies luminous enough to be included in the
sample out to $z=0.2$ ($M_{r^\ast}<-21.7$). We have quantified the
dependence on redshift with a linear regression along $z$. }
\end{figure}

\begin{figure}
\figurenum{\fignum}
\plotone{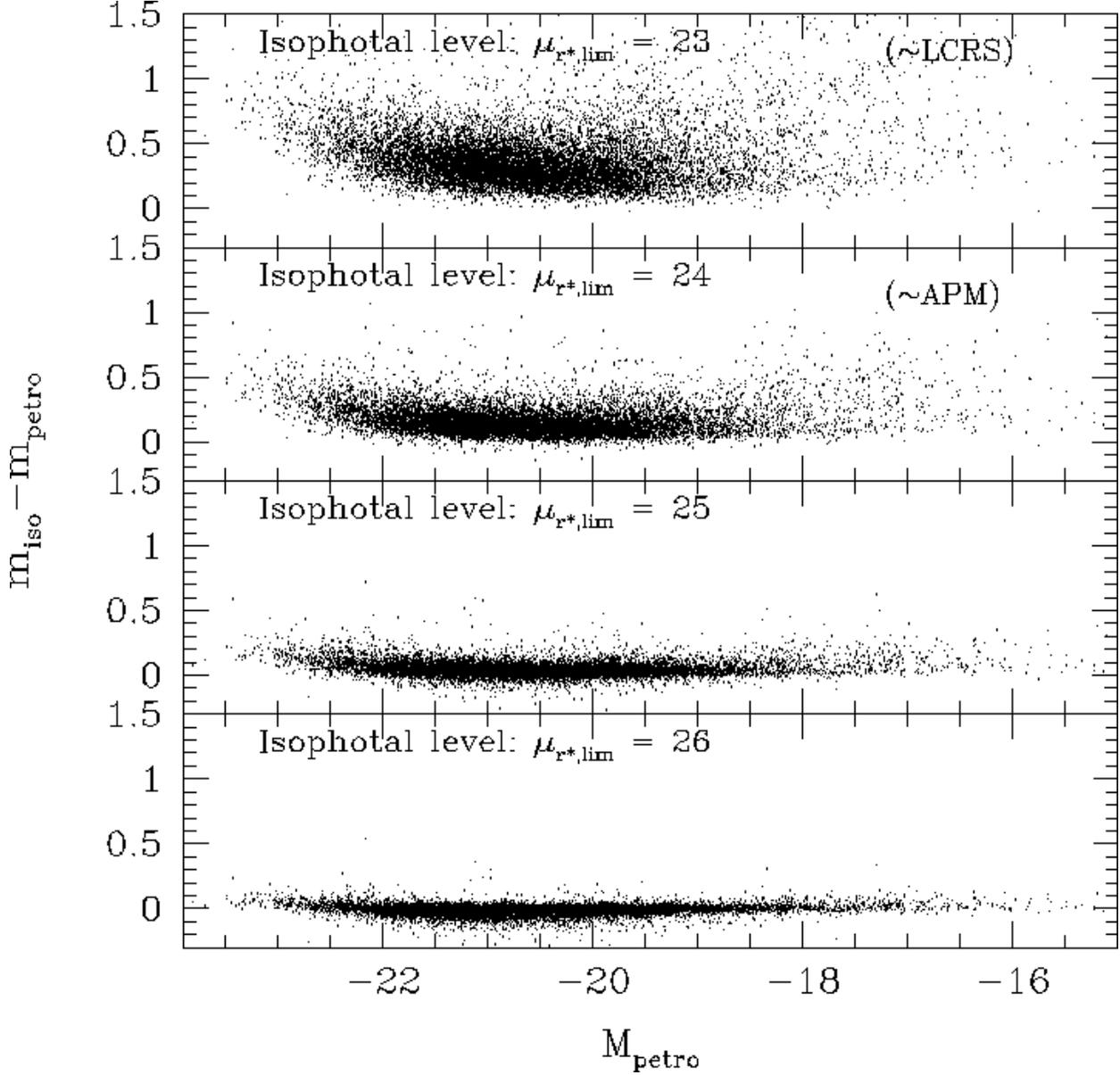}
\caption{\label{isomag_M} 
Comparison of isophotal and Petrosian magnitudes as a function of
absolute magnitude in the $r^\ast$ band for several choices of the
isophotal limit.  We show all galaxies with $r^\ast<17.6$.  The most
luminous galaxies have high $m_{\mathrm{iso}}-m_{\mathrm{petro}}$
because they tend to be further away, and thus suffer relatively more
surface brightness dimming. The least luminous galaxies have high
$m_{\mathrm{iso}}-m_{\mathrm{petro}}$ because they tend to be low
surface brightness.}
\end{figure}

Before selecting galaxies according to our newly defined isophotal
magnitudes and calculating a luminosity function, let us first
directly examine the relationship between the SDSS Petrosian
magnitudes and isophotal magnitudes.  If
$m_{\mathrm{iso}}-m_{\mathrm{petro}}$ were the same for all galaxies,
the luminosity scale would differ in the zero-point, but there would
be no other change in the shape of the luminosity function.  However,
$m_{\mathrm{iso}}-m_{\mathrm{petro}}$ depends on both surface
brightness (because low surface brightness objects hide a higher
fraction of their light below any given isophotal level) and redshift
(because of cosmological surface brightness dimming); thus the
resulting luminosity function is strongly affected.  To show the
redshift dependence, Figure \ref{isomag_z_vol} shows
$m_{\mathrm{iso}}-m_{\mathrm{petro}}$ for a volume-limited sample of
galaxies to $z=0.2$ (roughly $M_{r^\ast}<-21.7$), for several choices
of the isophotal limit. In each panel we show a linear regression fit,
showing clearly that the isophotal flux is not a constant fraction of
the Petrosian flux as a function of redshift.  Figure \ref{isomag_M}
shows $m_{\mathrm{iso}}-m_{\mathrm{petro}}$ as a function of absolute
magnitude. For luminous galaxies,
$m_{\mathrm{iso}}-m_{\mathrm{petro}}$ is large because most such
galaxies are near the edges of the survey (because more volume is
there) and thus suffer considerable surface brightness dimming. For
underluminous galaxies, $m_{\mathrm{iso}}-m_{\mathrm{petro}}$ is again
large, now because (as dictated by the correlation of luminosity and
surface brightness found in Section \ref{sbdep}) many of these objects
are low surface brightness.  The redshift and luminosity dependence
persists even at isophotal limits so faint that the typical redshift
zero isophotal magnitude is {\it brighter} than the same galaxy's
Petrosian magnitude. For isophotal limits more typical of other
surveys, such as $\mu_{r^\ast,\mathrm{iso}}=23$, which is comparable
to that of the LCRS, or at $\mu_{r^\ast,\mathrm{iso}}=24$, which under
the assumption that $\avg{b_j-r^\ast} \approx 1$ is comparable to the
2dFGRS, the trend with redshift is quite strong.
In Section \ref{others}, we will perform a much more careful
comparison of our results to those of the LCRS and the 2dFGRS.

\begin{figure}
\figurenum{\fignum}
\plotone{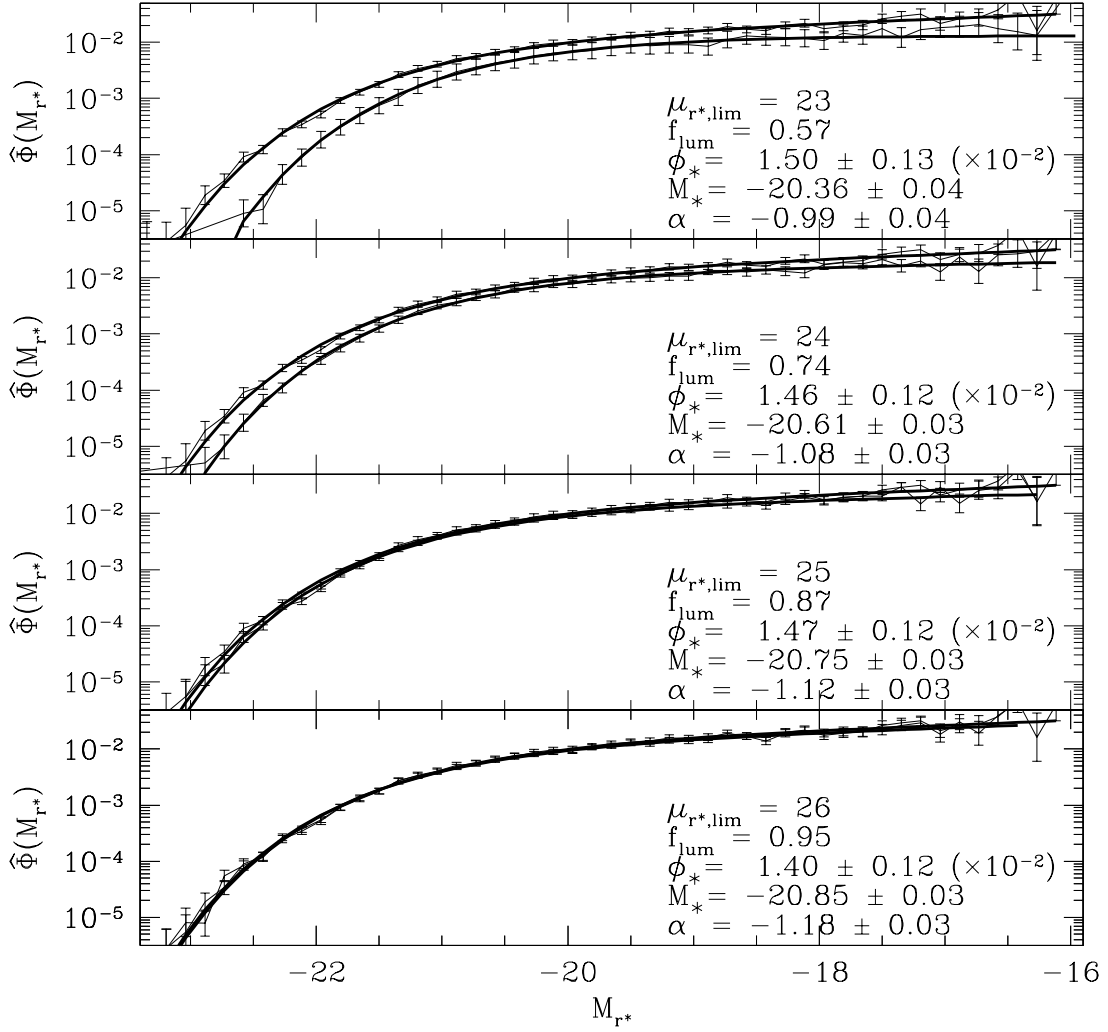}
\caption{\label{phi.iso} Fitted $r^\ast$ luminosity function for Petrosian
magnitudes (upper curve in each panel, same as Figure \ref{lf_full_r})
compared to the luminosity function using isophotal magnitudes (lower
curve in each panel). Each panel corresponds to a different choice of
limiting isophotal surface-brightness $\mu_{r^\ast,\mathrm{lim}}$, as
labeled. The Schechter function parameters for each choice of
isophotal limit are listed in each panel, as well as the fraction
$f_{\mathrm{lum}}$ of the total Petrosian luminosity that is detected
in each isophotal sample.}
\end{figure}

Figure \ref{phi.iso} compares the luminosity functions derived using
the isophotal and the Petrosian samples in the $r^\ast$ band. Clearly
there is a considerable difference between the estimates, which is not
attributable to a simple constant offset between isophotal and
Petrosian magnitudes. Instead, the isophotal magnitude estimate
differs from the Petrosian magnitude estimate at the low and high
luminosity ends. 
Figure \ref{phi.iso} also lists $f_{\mathrm{lum}}$, the fraction of
the integrated luminosity density in the Petrosian sample that is
recovered in each isophotal sample.  From these results, we infer that
surveys with shallower isophotal limits could be missing considerable
amounts of luminosity density.  We will compare our results more
directly to other surveys (and in the appropriate bands) in Section
\ref{others} below.

There are two reasons that isophotal samples can underestimate the
luminosity density when the isophotal limits are too bright. First,
isophotal magnitudes with bright limits measure less flux for each
galaxy, as revealed by Figure \ref{isomag_M}. Second, because the
fraction of the total flux measured by an isophotal magnitude
decreases with redshift from cosmological surface brightness dimming,
using the standard formula for the distance modulus overestimates the
effective volume of each galaxy (\citealt{dalcanton98a}). The dominant
effect in Figure \ref{phi.iso} is the first, that less light is
measured. To show this, we recalculate luminosity functions using a
simple $1/V_{\mathrm{max}}$ estimator ({\it e.g.},
\citealt{binggeli88a}). We first use the Petrosian magnitudes both to
determine the absolute magnitude of the galaxy and to apply the flux
limits (and thus also to determine $V_{\mathrm{max}}$). Then we use
the isophotal magnitudes to do both, finding the expected reduction in
luminosity density (about 40\% in the case of
$\mu_{r',\mathrm{lim}}=23$). We then calculate the luminosity function
again, determining $M_{r^\ast}$ from the isophotal magnitude but
applying the flux limits to the Petrosian magnitude. In this case, the
reduction of the luminosity density is nearly as great, around 35\%.
Finally, we determine $M_{r^\ast}$ from the Petrosian magnitude, but
apply the flux limits to the isophotal magnitude, thus isolating the
effect of an incorrect $V_{\mathrm{max}}$ determination. In the case
that $\mu_{r',\mathrm{lim}}=23$, this procedure reduced the luminosity
density only by 10\%. These results show that it is the missing light
beyond the isophotes that dominates the reduction in the luminosity
density, rather than an overestimate of the maximum
volume. Presumably, this result is related to that of
\citet{dalcanton98a}, that the systematic effects associated with
isophotal magnitudes have only a moderate effect on the estimated
$V/V_{\mathrm{max}}$ distribution. The error in $V_{\mathrm{max}}$
estimated with the standard distance modulus may be large in some
individual cases, but most galaxies enter the sample at a redshift
close to the largest one at which it is possible for them to, because
that is where most of the volume is. The additional dimming out to the
galaxy's limiting redshift is therefore small in most cases.

\section{Comparison to Other Surveys}
\label{others}

As we have seen above, the choice of isophotal magnitudes can greatly
affect the results for the galaxy luminosity function, confirming the
results of \citet{dalcanton98a}. Thus, in order to show that the
results of the SDSS are consistent with those of other surveys, we
must reanalyze the SDSS data in the same ways that previous surveys
analyzed their data.  The SDSS provides the ability to do so (for
surveys of comparable depth) in a way that no other large-scale survey
has before. Using the five-band color information and the color
conversions for galaxies provided by \citet{fukugita95a}, one can
convert the SDSS magnitudes into the appropriate band for almost any
desired optical survey. We convert to the LCRS $R$ band and to $b_j$,
and compare to the LCRS and 2dFGRS luminosity functions, finding
significantly more luminosity density in both cases.  Using the
measured photometric properties, one can reconstruct an isophotal
magnitude similar to that of any given survey (albeit with azimuthally
averaged light profiles). When we do so for the LCRS and for 2dFGRS,
we find our results are consistent with theirs. These results
demonstrate that the LCRS and 2dFGRS are missing significant fractions
of the light in the universe because their measurements of galaxy
magnitudes are based on fairly bright isophotes.

\subsection{Comparison to the LCRS}

\begin{figure}
\figurenum{\fignum}
\plotone{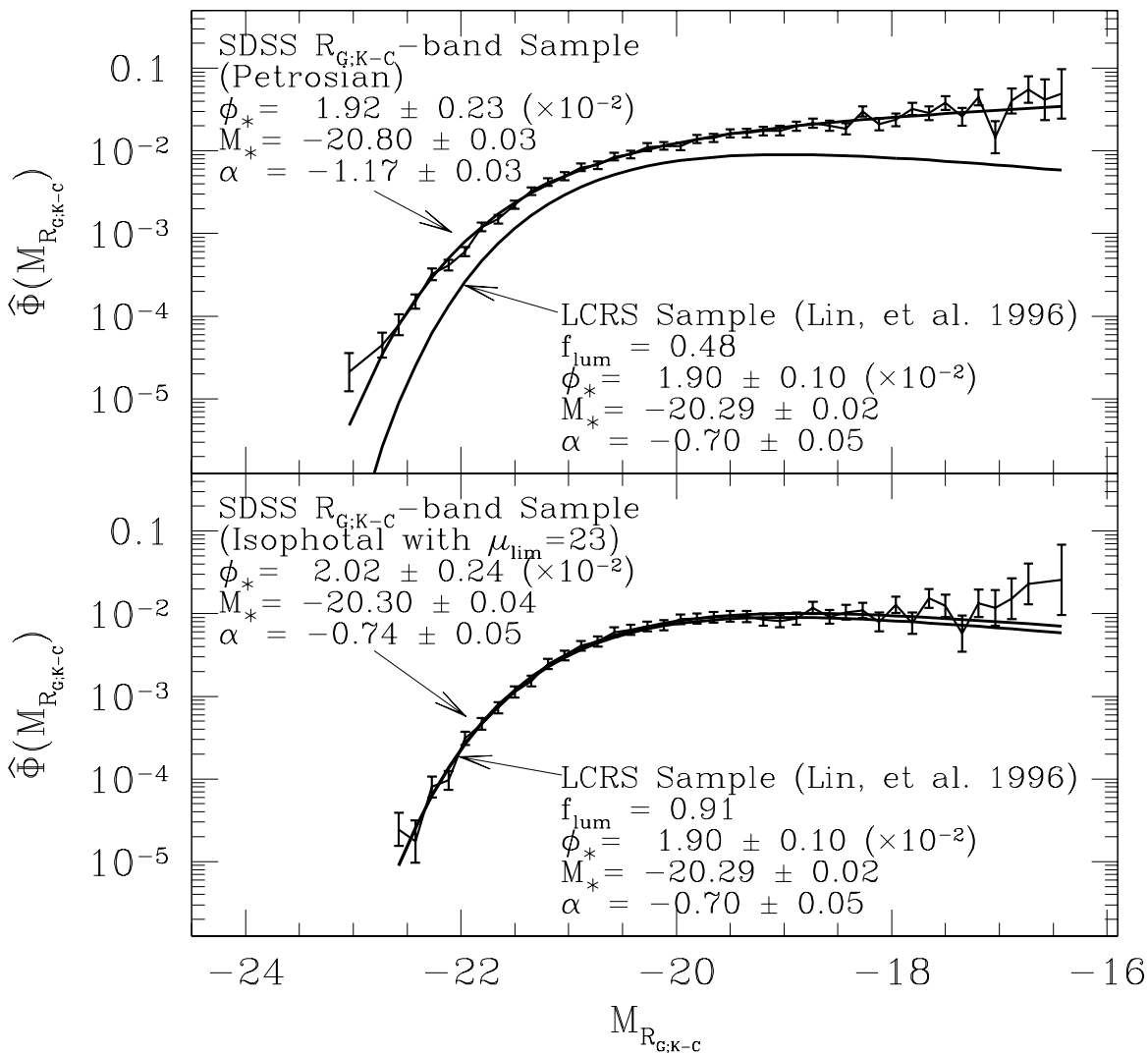}
\caption{\label{lf_lcrs} $R$-band luminosity function in the SDSS
compared to that in the LCRS calculated by \citet{lin96a}. Top panel
shows the luminosity function calculated from the SDSS using Petrosian
magnitudes. The ratio of the $R$-band luminosity density in the LCRS
to that in the SDSS is $f_{\mathrm{lum}}=0.48$. The bottom panel shows
the results of analyzing the SDSS data in the same manner as the LCRS,
using isophotal magnitudes limited at $\mu_{R,\mathrm{lim}} = 23$
and applying a central magnitude cut similar to that of the LCRS, as
described in the text. This sample is considerably more similar to the
results of the LCRS itself. The remaining discrepancies may be due to
our use of a slightly deeper isophotal magnitude than that of the
LCRS, the axisymmetric nature of our isophotal magnitudes, or the
slightly different definition of our central magnitude cut.}
\end{figure}

The LCRS photometry is expressed in its own variant of the $R$ band, a
hybrid of Gunn $r$ and the Kron-Cousins $R$ band, which we will denote
$R_{\mathrm{GKC}}$ following \citet{shectman96a}. This band can be related to
SDSS magnitudes as follows (\citealt{fukugita95a};
\citealt{shectman96a}):
\begin{equation}
R_{\mathrm{GKC}}=r^\ast-0.05-0.089(g^\ast-r^\ast).
\end{equation}
We use this transformation to calculate $R_{\mathrm{GKC}}$-band
Petrosian magnitudes for all SDSS galaxies, corrected for Galactic
extinction. Using the method described in Section \ref{method} for
defining complete samples in $u^\ast$, $g^\ast$, $i^\ast$ and
$z^\ast$, we define a complete sample in $R_{\mathrm{GKC}}$ by
setting a flux limit of $m_{\mathrm{lim}} = 17.40$. For consistency
with \citet{lin96a}, who calculate the $R_{\mathrm{GKC}}$-band
luminosity function in the LCRS, we use an Einstein-de Sitter universe
in this case.  The results are listed in Table
\ref{Rbj_table}. The top panel of Figure \ref{lf_lcrs} shows the
$R$-band luminosity function determined from the SDSS in this way and
compares our result to that of \citet{lin96a}. The two luminosity
functions diverge at low and high luminosities for the reasons
discussed in Section \ref{isopetro}: isophotal magnitudes
underestimate the luminosities of low luminosity galaxies because of
their low intrinsic surface brightnesses, and of high luminosity
galaxies because of cosmological surface brightness dimming. The ratio
of the $R$-band luminosity density found by \citet{lin96a} to that
found here is $f_{\mathrm{lum}} = 0.48$. That is, there is (at least)
twice as much $R_{\mathrm{GKC}}$-band light in the universe than
found by the LCRS. Table
\ref{Rbj_table} lists these results, along with the conversions of the
luminosity densities to solar units (assuming
$M_{R_{\mathrm{GKC}}\odot}=4.52$, using the results of
\citealt{binney98a} and the relation 
$R_{\mathrm{GKC}} = R_{\mathrm{KC}} + 0.1$ from
\citealt{shectman96a}; these choices are consistent with the results
quoted by \citealt{lin96a}).

Let us now analyze the SDSS images using the same method as used in
the LCRS. First, we use isophotal magnitudes limited at
$\mu_{R_{\mathrm{GKC}},\mathrm{lim}}=23$; this number approximately
corresponds to the statement of \citet{shectman96a} that the LCRS
isophotal magnitudes are limited at 15\% of the sky brightness.
Second,
\citet{shectman96a} exclude galaxies of low ``central surface
brightness,'' as measured by the magnitude within a fixed angular
aperture about the size of a fiber. Similarly, we apply a ``central
surface brightness'' cut based on the ``fiber magnitude,'' an aperture
magnitude with a $3''$ diameter.  Following \citet{shectman96a}, we
define the fiber magnitude cut to be:
\begin{equation}
\label{mfiber}
m_{R_{\mathrm{GKC}},\mathrm{fiber}} < 18.85 -
0.5(17.7-m_{R_{\mathrm{GKC}}}),
\end{equation}
Finally, for consistency we calculate the $K$-corrections using the
formula $K(z)=2.5\log_{10}(1+z)$, as \citet{lin96a} did (although this
makes rather little difference in our results). The bottom panel of
Figure \ref{lf_lcrs} shows the comparison of the resulting luminosity
function to that of \citet{lin96a}. They are largely consistent,
except that the SDSS sample has a slightly brighter $M_\ast$; this
difference may persist because our isophotal limit is not precisely
what LCRS used. Nevertheless, the general consistency shows that the
extra flux detected in the SDSS in the top panel is most likely real,
and that the missing light in the LCRS is due to having shallow
isophotes and excluding galaxies with low central surface
brightnesses. To understand how much is due to the central surface
brightness cut of Equation (\ref{mfiber}), we list results in Table
\ref{Rbj_table} for the case in which we do not apply this cut; this
test indicates that the central magnitude cut is responsible for a
substantial portion of the underestimate in the luminosity density.

\subsection{Comparison to the 2dFGRS}

\begin{figure}
\figurenum{\fignum}
\plotone{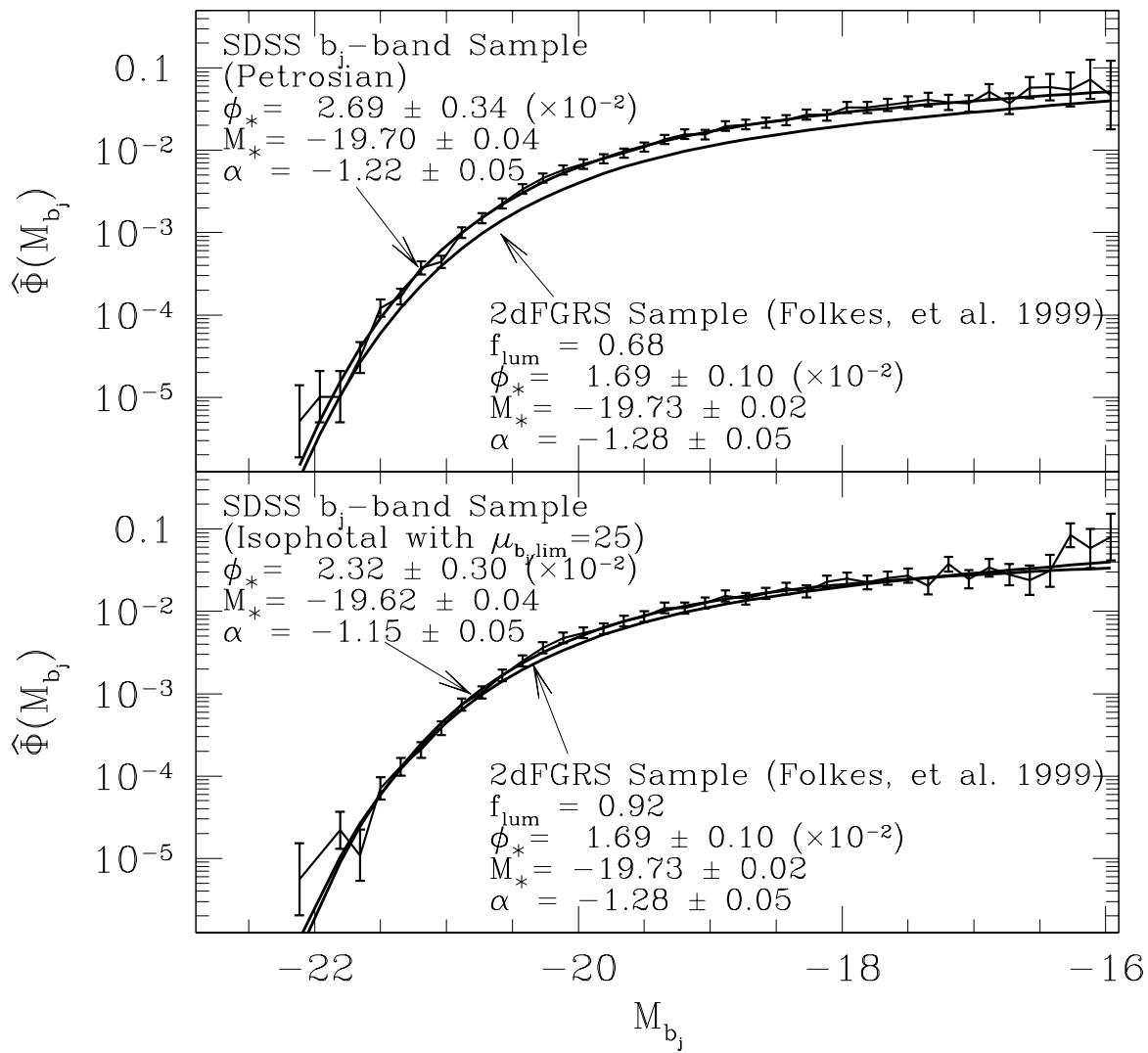}
\caption{\label{lf_apm} Same as Figure \ref{lf_lcrs}, this time
comparing the SDSS $b_j$ luminosity function to that of the 2dFGRS as
calculated by \citet{folkes99a}. The top panel shows that the SDSS
sample based on Petrosian magnitudes has about 1.4 times the
luminosity density of the 2dFGRS luminosity function. The bottom panel
shows that if we use isophotal magnitudes with the same isophotal
limit that the 2dFGRS uses, and ``correct'' these magnitudes in the
same manner as they do, we obtain nearly the same results. These
results indicate that the 2dFGRS misses a significant amount of
luminosity density due to its relatively shallow isophotal limits,
notwithstanding the efforts to correct the isophotal magnitudes
assuming a universal Gaussian profile for galaxies.}
\end{figure}

The 2dFGRS sample is based on APM plates (\citealt{maddox90a}) and
uses the $b_j$ band.  Here we relate the SDSS photometry to $b_j$
using the relation
\begin{equation}
b_j = g^\ast + 0.14 + 0.088(g^\ast-r^\ast),
\end{equation}
determined using the results of \citet{fukugita95a} and the galaxy
color relation $b_j = B - 0.35 (B-V)$ from \citet{metcalfe95a}.  We
follow the same procedure as for the LCRS in the previous section to
create a sample limited at $m_{b_j,{\mathrm{lim}}} = 17.8$. Again, for
consistency with \citet{folkes99a}, we assume an Einstein-de Sitter
universe. While our results are much closer to those of 2dFGRS than
they were to the LCRS, we still measure about 1.4 times the luminosity
density that \citet{folkes99a} do, as we show in Figure \ref{lf_apm}.
This difference is significant, given that the stated errors in the
$b_j$ luminosity density in the 2dFGRS are about 5\%, and the errors
for the SDSS are about 15\%. Table \ref{Rbj_table} lists these
results, now using $M_{b_j\odot}=5.30$ to convert to solar units
(determined using the stellar color relation $b_j=B-0.28(B-V)$ from
\citealt{blair82a} and the results of \citealt{binney98a}).

Let us analyze the SDSS images using the methods used by 2dFGRS.  APM
magnitudes are isophotally limited at $\mu_{b_j} = 25$ mag per square
arcsecond, and then corrected to total magnitudes by assuming that the
galaxy profile is a circularly symmetric Gaussian
(\citealt{maddox90b}).  After determining these ``APM-like''
magnitudes for SDSS galaxies, and applying a flux limit of
$m_{b_j,{\mathrm{lim}}}=17.8$, we again calculate the luminosity
function. This flux limit is considerably shallower than that of the
2dFGRS. Thus, the mean redshift is lower, and to accurately match the
isophotal limits would require a slightly brighter isophotal limit; we
have instead decided to err on the side of including a little more
flux.  The result is shown in the bottom panel of Figure \ref{lf_apm}.
The SDSS luminosity function determined in this manner is nearly
identical to the 2dFGRS luminosity function.  (Although the Schechter
parameters are rather different, it is clear from the figure that the
values are nearly degenerate).  Given our estimated errors and the
fact that our $b_j$ sample is considerably shallower than that of the
2dFGRS, this agreement of the two results is somewhat surprising.
This result indicates that the 40\% difference in luminosity density
between the SDSS and 2dFGRS is due to the fact that the APM isophotal
limits exclude a significant amount of light in the outer regions of
galaxies, even when corrected. For comparison, we also give results in
Table \ref{Rbj_table} for the case of $b_j=25$ isophotal magnitudes,
with no correction to ``total'' magnitudes based on the Gaussian
model. These results show that while the APM corrections work in the
right direction, the Gaussian model is an insufficient description of
observed galaxy profiles for extrapolation to ``total'' magnitudes.

\section{Conclusions}
\label{conclusions}

We have presented the luminosity function and luminosity density of
galaxies in five optical bandpasses ($u^\ast$, $g^\ast$, $r^\ast$,
$i^\ast$ and $z^\ast$) using $11,275$ galaxies in SDSS commissioning
data. The Schechter function appears to be a good fit in all cases,
with a low-luminosity slope that is remarkably similar in all
bandpasses. On the other hand, the slope $\alpha$ is sensitive to
galaxy surface brightness, color, and morphology. The redshift
distribution of the galaxies is generally consistent with our derived
luminosity function, although large-scale structure is still evident
in our sample.

We have demonstrated the correlation between $r^\ast$ luminosity and
galaxy surface brightness, color, and morphology. In particular, we
have shown that luminous galaxies tend to be higher surface
brightness, redder, and more concentrated than less luminous galaxies,
in accordance with previous results. We find that galaxies obey a
strong magnitude-surface brightness relation. There is a high
characteristic half-light surface brightness for luminous galaxies; at
the same time, low-luminosity galaxies show a broad surface brightness
distribution, with many low surface brightness objects. As described
above, refinements of the photometric analysis beyond the standard
software used for the survey will improve our understanding of these
correlations. For example, a better understanding of the effects of
seeing, inclination, and internal dust extinction will improve our
measures of surface brightness and morphology. Improvements in our
estimates of $K$-corrections, which will be possible using both the
spectroscopy and the five-band photometry of the SDSS, will improve
our understanding of the distribution of galaxy colors and allow us to
estimate the stellar mass function of galaxies.  Work in progress on
the spectral classification of SDSS galaxies ({\it e.g.},
\citealt{castander01a}) is showing the same trends of galaxy type with
luminosity found here.

We have shown that a considerable fraction of the luminosity density
in the universe has been missed by previous samples that were based on
shallower imaging. We measure 2 times the $R_{\mathrm{GKC}}$-band
luminosity density found in the LCRS and 1.4 times the $b_j$-band
luminosity density found in the 2dFGRS. Work in progress, which
directly compares the measured magnitudes of galaxies in these surveys
to SDSS magnitudes in the regions where the surveys overlap, will test
whether these differences are indeed reasonable. While large-scale
structure could still have some effect on results from these surveys,
our internal tests show that the differences of results are almost
certainly due to our adoption of Petrosian magnitudes in preference to
isophotal magnitudes. In particular, we can reproduce the LCRS and
2dFGRS results if we mimic their photometry and selection effects. The
high luminosity density found by the SDSS implies that the cosmic
density of stellar matter $\Omega_\ast$ is higher than previously
thought. If this is the case, it may relieve some of the strain on
galaxy formation models, which typically predict a higher density of
stellar matter than previous observational estimates
(\citealt{katz96a}; \citealt{pearce99a}; \citealt{granato00a}).

Our estimates of the luminosity density, given in Tables
\ref{lf_table} and \ref{Rbj_table}, are almost certainly still
underestimates. First, for the case of de Vaucouleurs profiles, we
know that up to 18\% of the galaxy light may be missed by the
Petrosian magnitudes as we define them. Including this light would
increase somewhat the total luminosity density found here (though only
by a few percent, because de Vaucouleurs profile galaxies constitute
only about 20\% of all galaxies). Second, we have shown that while the
explicit surface brightness limit $\mu_{1/2,\mathrm{lim}} = 23.5$ of
the commissioning data exclude only 0.5\% of galaxies, these galaxies
very likely represent a larger fraction of the luminosity density than
that; for example, only 1.5\% of galaxies are in the range 
$22.5 < \mu_{1/2,r^\ast} < 23.5$, but such galaxies are responsible
for $\sim$ 5\% of the luminosity in the universe. The deeper 
surface-brightness limit of the main SDSS survey will thus provide 
a more complete estimate of the luminosity density in the universe.
Third, our estimates assume a constant low luminosity slope
$\alpha$. An upturn in the luminosity function at absolute magnitudes
less luminous than $M=-16$, suggested by some studies
(\citealt{loveday97a}; \citealt{phillipps98a}), would mean there was
yet more luminosity not included in this analysis. A final
possibility, related to the previous one, is that a significant
fraction of stars exist outside of galactic environments, having
perhaps been tidally stripped from their galaxy of birth. As more SDSS
data is obtained and we perform more sophisticated analyses of the
various selection effects in the survey, we will be able to address
many of these problems.

Tables of the non-parametric fits to the luminosity function in each
band and the joint relationships between luminosity and surface
brightness, color, and concentration, are available on the World Wide
Web\footnote{{\tt
http://www-astro-theory.fnal.gov/Personal/blanton/sdss-lf/}}.  We
expect that these results and the parametric fits in Table
\ref{lf_table}, which represent a straightforward analysis
of only a small fraction (one percent) of the expected SDSS sample,
will already provide interesting new constraints on theories of galaxy
formation. In addition, this characterization of the luminosities,
surface brightnesses, colors, and morphologies of local galaxies
provides a solid baseline for interpreting the evolution of galaxies
observed at higher redshift. Future work with the SDSS will allow us
to explore the correlations between all of these properties and to
describe fully the population of galaxies in the local universe.

\acknowledgments

We would like to thank Ravi Sheth, Ned Wright, Idit Zehavi, and an
anonymous referee for comments, advice, and corrections. MB is
supported by the DOE and NASA grant NAG 5-7092 at Fermilab, and is
grateful for the hospitality of the Department of Physics and
Astronomy at the State University of New York at Stony Brook, who
kindly provided computing facilities on his frequent visits there.
DJE was supported by NASA through Hubble Fellowship grant
\#HF-01118.01-99A from the Space Telescope Science Institute, which is
operated by the Association of Universities for Research in Astronomy,
Inc, under NASA contract NAS5-26555. MAS acknowledges the support of
NSF grant AST-0071091.

The Sloan Digital Sky Survey (SDSS) is a joint project of The
University of Chicago, Fermilab, the Institute for Advanced Study, the
Japan Participation Group, the Johns Hopkins University, the
Max-Planck-Institute for Astronomy, New Mexico State University,
Princeton University, the United States Naval Observatory, and the
University of Washington. Apache Point Observatory, site of the SDSS
telescopes, is operated by the Astrophysical Research Consortium
(ARC).  Funding for the project has been provided by the Alfred
P.~Sloan Foundation, the SDSS member institutions, the National
Aeronautics and Space Administration, the National Science Foundation,
the U.~S.~Department of Energy, Monbusho, and the Max Planck
Society. The SDSS Web site is http://www.sdss.org/.

\newpage

\clearpage
 
\begin{deluxetable}{cccr}
\tablewidth{0pt}
\tablecolumns{4}
\tablecaption{\label{lf_limits} Limits Applied to Sample in Each SDSS
Band}
\tablehead{ Band & Flux Limits & Redshift Limits (km s$^{-1}$) & Number of Galaxies}
\startdata
$u^\ast$ & $14.50 < m < 18.40$ & $5,000 < cz < 30,000$ & 1,679\cr
$g^\ast$ & $14.50 < m < 17.65$ & $5,000 < cz < 50,000$ & 4,684\cr
$r^\ast$ & $14.50 < m < 17.60$ & $5,000 < cz < 60,000$ & 11,275\cr
$i^\ast$ & $14.50 < m < 16.90$ & $5,000 < cz < 60,000$ & 7,441\cr
$z^\ast$ & $14.50 < m < 16.50$ & $5,000 < cz < 60,000$ & 6,090\cr
\enddata
\end{deluxetable}

\begin{deluxetable}{cccccccccc}
\tablewidth{0pt}
\tablecolumns{10}
\tablecaption{\label{lf_table} Parameters of Fits to Luminosity
Function in SDSS}
\tablecomments{Luminosity function in various bands assuming the three
cosmological models considered here, with $H_0=100$ km s$^{-1}$. The
Schechter function is characterized by the three parameters
$\phi_\ast$ (in units of $10^{-2}$ $h^3$ Mpc$^{-3}$), $M_\ast$, and
$\alpha$, whose values and errors in each case are listed here. Also
listed is the covariance between the errors of $M_\ast$ and $\alpha$,
determined from the Monte Carlo error determination.  $j$ is the
luminosity density as determined by integrating the Schechter function
over all luminosities (again in units of $h^3$ Mpc$^{-3}$). We list
$j$ both in absolute magnitudes and in solar luminosities (determined
as described in the text). The error bars on the value of $j$ in solar
luminosities include a 7\% contribution due to calibration and color
transformation uncertainties. $f_{\mathrm{np}}$ is the fraction of the
luminosity in the extrapolated Schechter function which is accounted
for in the non-parametric fit; that is, this is the fraction of the
inferred light which comes from galaxies with luminosities which are
observed in the sample. Naturally, it is lowest when the slope
$\alpha$ at low luminosities is highest.}
\tablehead{ $\Omega_m$ & $\Omega_\Lambda$ & Band & $\phi_\ast$ &
$M_\ast - 5 \log_{10} h$ & 
$\alpha$ & $r_{M_{\ast},\alpha}$ & $j$ (mags) & $j$ ($h$ $10^{8} L_{\odot}$) &
$f_{\mathrm{np}}$ \\
 & &  & ($10^{-2}$ $h^3$ Mpc$^{-3}$) & & & &
& }
\startdata
0.3 & 0.7 & $u^\ast$ &
$4.00 \pm 0.90$ & $-18.34 \pm 0.08$ & $-1.35 \pm 0.09$ & 0.80 & 
$-15.21 \pm 0.26$ & $4.35 \pm 1.08$ & 0.85 \cr
 &  & $g^\ast$ &
$2.06 \pm 0.23$ & $-20.04 \pm 0.04$ & $-1.26 \pm 0.05$ & 0.79 & 
$-16.05 \pm 0.13$ & $2.81 \pm 0.38$ & 0.94 \cr
 &  & $r^\ast$ &
$1.46 \pm 0.12$ & $-20.83 \pm 0.03$ & $-1.20 \pm 0.03$ & 0.78 & 
$-16.41 \pm 0.09$ & $2.58 \pm 0.28$ & 1.00 \cr
 &  & $i^\ast$ &
$1.28 \pm 0.11$ & $-21.26 \pm 0.04$ & $-1.25 \pm 0.04$ & 0.77 & 
$-16.74 \pm 0.10$ & $3.19 \pm 0.37$ & 0.98 \cr
 &  & $z^\ast$ &
$1.27 \pm 0.11$ & $-21.55 \pm 0.04$ & $-1.24 \pm 0.05$ & 0.74 & 
$-17.02 \pm 0.11$ & $3.99 \pm 0.48$ & 0.97 \cr
0.3 & 0.0 & $u^\ast$ &
$4.48 \pm 1.06$ & $-18.27 \pm 0.07$ & $-1.32 \pm 0.09$ & 0.78 & 
$-15.20 \pm 0.26$ & $4.34 \pm 1.07$ & 0.84 \cr
 &  & $g^\ast$ &
$2.48 \pm 0.33$ & $-19.97 \pm 0.05$ & $-1.23 \pm 0.05$ & 0.83 & 
$-16.16 \pm 0.15$ & $3.09 \pm 0.47$ & 0.98 \cr
 &  & $r^\ast$ &
$1.71 \pm 0.15$ & $-20.74 \pm 0.03$ & $-1.20 \pm 0.03$ & 0.74 & 
$-16.49 \pm 0.09$ & $2.78 \pm 0.31$ & 0.96 \cr
 &  & $i^\ast$ &
$1.47 \pm 0.13$ & $-21.17 \pm 0.04$ & $-1.23 \pm 0.04$ & 0.77 & 
$-16.78 \pm 0.10$ & $3.31 \pm 0.39$ & 0.96 \cr
 &  & $z^\ast$ &
$1.45 \pm 0.14$ & $-21.46 \pm 0.04$ & $-1.20 \pm 0.05$ & 0.80 & 
$-17.02 \pm 0.11$ & $3.99 \pm 0.48$ & 1.00 \cr
1.0 & 0.0 & $u^\ast$ &
$4.68 \pm 1.13$ & $-18.24 \pm 0.07$ & $-1.31 \pm 0.09$ & 0.82 & 
$-15.20 \pm 0.26$ & $4.34 \pm 1.10$ & 0.84 \cr
 &  & $g^\ast$ &
$2.42 \pm 0.30$ & $-19.92 \pm 0.04$ & $-1.22 \pm 0.05$ & 0.80 & 
$-16.07 \pm 0.13$ & $2.87 \pm 0.40$ & 0.97 \cr
 &  & $r^\ast$ &
$1.87 \pm 0.18$ & $-20.67 \pm 0.03$ & $-1.15 \pm 0.03$ & 0.84 & 
$-16.47 \pm 0.10$ & $2.74 \pm 0.31$ & 1.00 \cr
 &  & $i^\ast$ &
$1.61 \pm 0.15$ & $-21.11 \pm 0.04$ & $-1.20 \pm 0.05$ & 0.78 & 
$-16.78 \pm 0.11$ & $3.32 \pm 0.40$ & 0.98 \cr
 &  & $z^\ast$ &
$1.62 \pm 0.16$ & $-21.39 \pm 0.04$ & $-1.19 \pm 0.05$ & 0.74 & 
$-17.08 \pm 0.11$ & $4.19 \pm 0.51$ & 0.96 \cr
\enddata
\end{deluxetable}


\begin{deluxetable}{ccccccc}
\tablewidth{0pt}
\tablecolumns{7}
\tablecaption{\label{Rbj_table} Comparison of SDSS Luminosity Function
to 2dFGRS and LCRS}
\tablecomments{
Same as Table \ref{lf_table}, for the LCRS $R_{\mathrm{G;K-C}}$-band
and the $b_j$ band. We include comparisons to the LCRS
(\citealt{lin96a}) and the 2dFGRS (\citealt{folkes99a}). }
\tablehead{ Band & Sample & $\phi_\ast$ & $M_\ast - 5\log_{10} h$ & 
$\alpha$ & $j$ (mags) & $j$ ($h$ $10^{8} L_{\odot}$) \\
 &  & ($10^{-2}$ $h^3$ Mpc$^{-3}$) & & & &
}
\tablenotetext{a}{Without the limits on $m_{\mathrm{fiber}}$ in
Equation (\ref{mfiber}).}
\tablenotetext{b}{With the limits on $m_{\mathrm{fiber}}$ in
Equation (\ref{mfiber}).}
\tablenotetext{c}{Without APM-like corrections to the isophotal
magnitudes.}
\tablenotetext{d}{With APM-like corrections to the isophotal
magnitudes.}
\startdata
$R_{\mathrm{G;K-C}}$ & SDSS Petrosian & 
$1.92 \pm 0.23$ & $-20.80 \pm 0.03$ & $-1.17 \pm 0.03$ & 
$-16.64 \pm 0.12$ & $2.91 \pm 0.38$ \cr
$$ & SDSS Isophotal\tablenotemark{a} & 
$2.13 \pm 0.26$ & $-20.37 \pm 0.04$ & $-0.94 \pm 0.05$ & 
$-16.15 \pm 0.12$ & $1.86 \pm 0.24$ \cr
$$ & SDSS Isophotal\tablenotemark{b} & 
$2.02 \pm 0.24$ & $-20.30 \pm 0.04$ & $-0.74 \pm 0.05$ & 
$-15.95 \pm 0.12$ & $1.54 \pm 0.20$ \cr
$$ & LCRS & 
$1.90 \pm 0.1$ & $-20.29 \pm 0.02$ & $-0.70 \pm 0.03$ & 
$-15.87 \pm 0.1$ & $1.4 \pm 0.1$ \cr
$b_j$ & SDSS Petrosian & 
$2.69 \pm 0.34$ & $-19.70 \pm 0.04$ & $-1.22 \pm 0.05$ & 
$-15.97 \pm 0.14$ & $3.21 \pm 0.46$ \cr
$$ & SDSS Isophotal\tablenotemark{c} & 
$2.17 \pm 0.28$ & $-19.55 \pm 0.05$ & $-1.12 \pm 0.06$ & 
$-15.47 \pm 0.14$ & $2.04 \pm 0.30$ \cr
$$ & SDSS Isophotal\tablenotemark{d} & 
$2.32 \pm 0.30$ & $-19.62 \pm 0.04$ & $-1.15 \pm 0.05$ & 
$-15.65 \pm 0.13$ & $2.39 \pm 0.34$ \cr
$$ & 2dFGRS & 
$1.69 \pm 0.17$ & $-19.73 \pm 0.06$ & $-1.28 \pm 0.05$ & 
$-15.56 \pm 0.05$ & $2.19 \pm 0.12$ \cr
\enddata
\end{deluxetable}

\clearpage
\clearpage

\setcounter{thefigs}{0}

\clearpage
\stepcounter{thefigs}

\clearpage
\stepcounter{thefigs}

\clearpage
\stepcounter{thefigs}

\clearpage
\stepcounter{thefigs}

\clearpage
\stepcounter{thefigs}

\clearpage
\stepcounter{thefigs}

\clearpage
\stepcounter{thefigs}

\clearpage
\stepcounter{thefigs}

\clearpage
\stepcounter{thefigs}

\clearpage
\stepcounter{thefigs}

\clearpage
\stepcounter{thefigs}

\clearpage
\stepcounter{thefigs}

\clearpage
\stepcounter{thefigs}

\clearpage
\stepcounter{thefigs}

\clearpage
\stepcounter{thefigs}

\clearpage
\stepcounter{thefigs}

\clearpage
\stepcounter{thefigs}

\clearpage
\stepcounter{thefigs}


\clearpage
\stepcounter{thefigs}

\clearpage
\stepcounter{thefigs}

\clearpage
\stepcounter{thefigs}

\clearpage
\stepcounter{thefigs}

\clearpage
\stepcounter{thefigs}

\end{document}